\documentclass[onecolumn,noshowpacs,nofootinbib,colorlinks,hyperindex,unicode,12pt]{revtex4}

\usepackage{float}
\usepackage{blkarray, bigstrut}
\usepackage{multirow}
\usepackage{arydshln}
\usepackage{amsfonts,amsmath,amssymb}
\usepackage{accents}
\usepackage{cancel}

\usepackage{mathptmx,mathrsfs}

\usepackage{enumitem}

\usepackage{ragged2e}

\makeatletter
\renewcommand*\env@matrix[1][*\c@MaxMatrixCols c]{%
  \hskip -\arraycolsep
  \let\@ifnextchar\new@ifnextchar
  \array{#1}}
\makeatother

\usepackage{hyperref,upgreek}
\usepackage{slashed}
\counterwithout{equation}{section}
\usepackage{tikz}
\usetikzlibrary{matrix}

\hypersetup{hidelinks,backref=true,pagebackref=true,hyperindex=true,colorlinks=true,breaklinks=true,urlcolor= blue}
\hypersetup{%
  colorlinks = true,
  linkcolor  = blue,
  citecolor = blue,
}
\usepackage{textgreek}
\usepackage{color,xcolor}

\newcommand\topstrut[1][1.2ex]{\setlength\bigstrutjot{#1}{\bigstrut[t]}}
\newcommand\botstrut[1][0.9ex]{\setlength\bigstrutjot{#1}{\bigstrut[b]}}

\newcommand{\orcidicon}{
	\begin{tikzpicture}
	\draw[lime, fill=lime] (0,0)
		circle [radius=0.16]
		node[white] {{\fontfamily{qag}\selectfont \tiny ID}};
	\draw[white, fill=white] (-0.0625,0.095)
		circle [radius=0.007];
	\end{tikzpicture}	\hspace{-2mm}
}
\newcommand\orcidg{{\href{https://orcid.org/0000-0002-7942-7941}{\orcidicon}}}
\newcommand\orcidNog{{\href{https://orcid.org/0000-0002-1827-1031}{\orcidicon}}}
\newcommand\orcidLG{{\href{https://orcid.org/0000-0002-6525-7857}{\orcidicon}}}
\newcommand\orcidBMP{{\href{https://orcid.org/0000-0002-2376-8253}{\orcidicon}}}
\begin{document}

\title{Path Integral Quantization of Generalized Stueckelberg Electrodynamics:
A Faddeev-Jackiw Approach}

\author{L. G. Caro\orcidLG{}}
\email{gabriel.caro@unesp.br}
\affiliation{São Paulo State University, Institute for Theoretical Physics, São Paulo, 01140-070, Brazil}
\author{G. B. de Gracia\orcidg{}}
\email{g.gracia@ufabc.edu.br}
\affiliation{Federal University of ABC, Center of Mathematics, Santo Andr\'e, 09210-580, Brazil}
\author{A. A. Nogueira\orcidNog{}}
\email{anderson.nogueira@unifal-mg.edu.br}
\affiliation{Federal University of Alfenas, Institute of Exact Sciences, Alfenas 37133-840, Brazil}
\author{B. M. Pimentel\orcidBMP{}}
\email{bruto.max@unesp.br}
\affiliation{São Paulo State University, Institute for Theoretical Physics, São Paulo, 01140-070, Brazil}

\begin{abstract}

\vspace{2.5ex}
\centering \textbf{Abstract}
\justify
We build {\color{blue} a} setup for path integral quantization through the Faddeev-Jackiw approach, extending it to include Grassmannian degrees of freedom, to be later implemented in a model of generalized electrodynamics that involves fourth-order derivatives in the components of a massive vector field being endowed with gauge freedom, {\color{blue}due to} an additional scalar field. Namely, the  generalized Stueckelberg electrodynamics. In the first instance, we work on the free case to gain some familiarity with the program and, subsequently, we add the interaction with fermionic matter fields to complete our aim. In addition to deriving the correct classical brackets for such a model, we get the full expression for the associated generating functional and its associated integration measure.
\vspace{2ex}\\

\end{abstract}

\maketitle
\date{}
\newpage

\section{Introductory Aspects}
\label{sc1}


\indent Considering the context in which the quantum equations of motion were further developed from classical dynamics, Dirac \cite{WdW} described them with a formal language using the correspondence principle. He discovered that there is a correspondence between classical dynamics in phase space (described by dynamic functions and Poisson brackets (PB)) and quantum dynamics in a Hilbert space (described by  self-adjoint operators and commutators). Some time later, Dirac made the first thorough analysis of constrained systems in a Hamiltonian framework, probably one of his main achievements being the classification of constraints into first and second class, since the former were recognized as generators of gauge transformations. Besides that, Dirac's work gave rise to the so-called Dirac brackets, which are a generalized version of the PB and the correct ones for that kind of system. This Dirac's approach was further extended to connect classical and quantum dynamics of constrained systems via canonical quantization, by the Dirac-Bergmann (DB) algorithm \cite{Dirac1, Bergmann}.

The need for a relativistic description of the interactions that take place in nature led to the development of a covariant language with gauge freedom \cite{Ut,Ut2}. However, this language involves more degrees of freedom than necessary, which implies the presence of constraints. Faddeev was the first to formulate the connections between classical and quantum descriptions of physical systems with constraints in a functional formalism, and Senjanovic later extended these ideas \cite{Fadd}. The quantization of a gauge theory is only possible in the physical degrees of freedom, resulting in a loss of explicit covariance in the equations. Nevertheless, in order to maintain explicit covariance at the quantum level, Faddeev and others developed a method that introduces non-physical ghost particles \cite{faddeev}.


Following the historical development of quantization procedures of constrained systems, the Faddeev-Jackiw (FJ) method revitalized the subject in the 1980s \cite{FJ}. This formalism takes a classical geometric approach based on the symplectic structure of the phase space, and it is only applied to Lagrangians  with linear dependence in the velocities. The method was originally based on the generalized Darboux's theorem for pre-symplectic forms, which are characteristic of phase space of constrained systems, to identify the constraints of the theory as well as the coordinates that describe the associated true degrees of freedom. In these coordinates, a reduced phase space is defined, which is equipped with a symplectic form whose inverse allows us to determine the modified Poisson brackets. In the majority of the cases of interest, it coincides with the Dirac brackets. In fact, many studies have explored the equivalence between the FJ and DB formalisms, with the equivalence being proved in some but not all cases \cite{Jairzinho1,Li2,Li3,Jairzinho2, Zambrano}.  It is worth mentioning that in this approach one does not have to go through the steps of Dirac's method \cite{Jackiwtears}.

Certainly, the FJ approach offers several advantages, such as not having to distinguish between types of constraints and bypassing the Dirac conjecture. This led to an increase in interest in the formalism, so that a few years later Barcelos-Neto and Wotzasek made a contribution to the method by proposing an alternative path that circumvents the use of Darboux's theorem and instead uses the emerging constraints to extend the phase space iteratively inserting Lagrange multipliers as new coordinates until obtaining a new set that allows defining a symplectic structure on it  \cite{BW,MW}. 

On the other hand, by studying the fundamental role of the Lagrangian in quantum mechanics, Dirac introduced the very first idea of what is known today as transition amplitude and its relation with the classical action \cite{Dirac2}. This idea was later taken by Feynman to develop its path integral formulation, in which the classical action plays a key role \cite{Feyn}. Subsequently, Schiwnger postulates its (variational) quantum action principle in the Heisenberg description
\cite{Schwinger,Nakani}. 
As we will see below, the functional integration formalism will trigger several lines of our research in the study of the quantization procedure, since, despite its primarily canonical nature, the FJ method has also been used in path integral quantization \cite{Li1,Toms}. 

Now, taking into account that the wide study of models that involve not only gauge freedom, but also massive vector fields, in the context of higher-order theories \cite{Ostro,Bop,mass1,mass2,mass3,mass4}, as well as the different treatments of them (such as the DB algorithm or the FJ symplectic analysis) have led to several investigations on the quantization procedure \cite{galvao,bufy,HJ,Reduced,andy}, it is plausible to complete these investigations and explore the subtleties of the FJ quantization approach in a new environment \cite{m2y}. Namely, the Generalized Stueckelberg (GS) electrodynamics. Finally, it is worth mentioning the fact that the Ostrogadsky instabilities can be overcome by quantum radiative corrections \cite{Russos, Merlim}.

This article is organized as follows: In Sec. \ref{sc2} we present a general background of the free GS model, subsequently, the FJ framework is detailed in Sec. \ref{sc3} to be later applied to the free GS model in Sec. \ref{sc4}. Thereupon, the path integral quantization in the FJ perspective and its application in the free GS model are presented in Sec. \ref{sc5}. Thereafter, looking to incorporate interactions to the previous model, in Sec. \ref{sc6} we extend the FJ theory to the case in which Grassmann-valued fields are present. With all these tools ready, we make in Sec. \ref{sc7} the FJ analysis and the corresponding path integral quantization of the GS electrodynamics. Finally, Sec. \ref{sc8} contains the epilogue of the investigations, with our conclusions and perspectives of future work.

\section{The Free GS Model}
\label{sc2}

\indent It is well known that the Podolsky model describes an ensemble of massive and massless generalized photons by means of an Ostrogadskian structure of fourth order in derivatives. This implies an ultraviolet improvement for the  case of an interacting theory \cite{Renorma,RenorDark}. However, even with this refinement, the model still presents infrared divergences. Consequently, one of the goals  of a suitable extension is to provide a low-energy regularization for the theory, which can be achieved by a usual Podolsky theory plus a mass term written with a Stueckelberg combination, responsible for keeping the gauge invariance. This combination of $U.V.$ and $I.R.$ mass terms can be obtained by a generalization of the Higgs mechanism \cite{m2y}. The analysis of the interacting version of this model for $T\neq 0$ in the context of the Debye screening discussion was performed in \cite{Deb}.

\indent The free case is described by the following Lagrangian density 
\begin{equation}
\mathcal{L}=\underbrace{-\frac{1}{4}F_{\mu \nu}F^{\mu \nu}}_{\mathcal{L}_{M}} + \underbrace{\frac{1}{2m^2}\partial^\mu F_{\rho\mu}\partial_\nu F^{\rho\nu}  }_{\mathcal{L}_{P}}+\underbrace{\frac{M^2}{2}\left(A_\mu + \dfrac{1}{m_s}\partial_\mu B\right)\left(A^\mu + \dfrac{1}{m_s}\partial^\mu B\right)}_{\mathcal{L}_{S}}\;.
\label{FPS}
\end{equation}
It is important to make a comment on the different contributions in the  full Lagrangian density $\mathcal{L}$ above. The first term corresponds to the well-known Maxwell Lagrangian density (describing the dynamics of the free electromagnetic field). On the other hand, the two first terms (i.e., $\mathcal{L}_{M}+\mathcal{L}_{P}$) correspond to the (free) Podolsky electrodynamics, which is a second-order theory of a real vector field. Finally, $\mathcal{L}_{M}+\mathcal{L}_{S}$ leads to the free Stueckelberg model, which describes a coupling between a massive real vector field and a real scalar field $B$, in order to maintain the gauge freedom. Considering the limit $m\to +\infty$, we get the Stueckelberg-Proca theory improving the I.R. sector; on the other hand, for $m\to +\infty $ and $M\to 0$, we recover the free Maxwell electrodynamics. Finally, for $M\to 0$, we get the free Podolsky model leading to U.V completion. Therefore, the complete model ensures improvements in both the U.V. and the I.R. regime.

It is easy to check that the total Lagrangian density $\mathcal{L}$ remains invariant under the following local gauge transformations
\begin{equation}
A_\mu \to A'_\mu=A_\mu+\partial_\mu \Lambda \qquad , \qquad  B\to B'=B -m_s\Lambda \;,
\label{gauge.tr}
\end{equation}    
with $\Lambda=\Lambda(x)$ completely arbitrary. Interestingly, in the next section, it is going to be proved that the pure gauge Stueckelberg field $B$ does not contribute to the physical degrees of freedom, as it should be.

\indent It is worth mentioning that if we use a suitable covariant gauge\footnote{The Lorenz-Podolsky-Stueckelberg gauge condition is given by $\left(1+\dfrac{\Box}{m^2}\right)\partial_\mu A ^\mu - \dfrac{M^2}{m_s}B=0$.}, it is possible to eliminate the harmonic Stueckelberg field and that way get the following equation of motion
\begin{equation}
    \left[\left(1+\frac{\Box}{m^2}\right)\Box+M^2\right]A_\mu=J_\mu     \;,
\end{equation}
after the introduction of an external source $J_\mu$ in the Lagrangian in \eqref{FPS} by adding a term $- J_\mu A^\mu$.

\indent For the particular case of a static charge at the origin, i. e. $J_\mu(x)=\delta_{\mu 0}\,q\,\delta^3(\vec x-\vec0)$, the induced vacuum average for the inter-particle potential becomes \cite{Deb}
\begin{equation}
A_0(\vec x)=\dfrac{m^2q}{4\pi r}\left(\dfrac{e^{-m_-r}-e^{-m_+r}}{m^2_+-m^2_-}\right)\;;
\end{equation}
with $r$ representing the modulus of spatial position $\vec x $, and the pole masses are defined as being
\begin{equation}
m_\pm=\sqrt{\frac{1}{2}\left(m^2\pm m\sqrt{m^2-4M^2}\right)}   \;.  
\end{equation}
It is interesting to note that, depending on the values of $m$ and $M$, a complex pole can be reached leading to a confining phase \cite{Zwan}. Additionally, one of the main properties of this model is to present three distinct phases for the potential $A_0$. A mixture of Yukawa-like terms for $M<\frac{m}{2}$, an exponential decaying one for $M=\frac{m}{2}$  (resembling the magnetic field behavior inside superconductors), and an oscillatory phase associated  with the regime $M>\frac{m}{2}$. Just for comparison, if a Proca-like pole $M$ is generated through thermal corrections for the usual interacting Podolsky model, it would be\footnote{With $\alpha$ being the fine structure constant and $T$, the temperature.} $M\sim \sqrt{\frac{4\pi \alpha}{3}}T$. Therefore, the non-trivial regime is reached at an incredibly large temperature of order $10^{23}\,K$.

\indent Although investigating several well-defined possibilities for extensions of physical theories is an interesting subject by itself, the changes must be constrained by experimental observations. For instance, the Podolsky mass is restricted to the range $m\geqslant 370\ GeV$, to be in accordance with electron-positron interaction phenomenology \cite{BufSot}. Besides, there is a very stringent bound on the Proca mass $M\leq 10^{-12}\ eV $ according to the analysis of samples from extragalactic radio pulsars \cite{JunJie}. It is worth noting that the investigation into the origin of the Podolsky mass remains an ongoing subject of current research \cite{BoninM}.

\indent The variety of interesting properties convinced us that such a model can furnish useful insights on various aspects of field theory, if investigated by several phase space approaches such as DB formalism \cite{galvao}, Hamilton-Jacobi analysis \cite{HJ} and the FJ framework \cite{bufy,andy}. This program was implemented for several field theories, revealing important complementary achievements.

\section{Standard FJ Formulation}
\label{sc3}

\indent Since our aim is to implement the Faddeev-Jackiw approach on the free GS model, we present its formal framework in the context of a field theory, following Barcelos-Neto and Wotzasek \cite{BW}. The starting point is a Lagrangian written in its canonical form, i.e., linear in the velocities
\begin{equation}
L[\xi,\dot{\xi}]=\int  \theta_I(x)\,\dot{\xi}^I(x)d^3x-H[\xi]\qquad , \qquad \dot{\xi}^I\overset{!}{=}  \partial_0\xi^I \;;
\label{can.lag}
\end{equation}
in which  $\xi^I$ denotes the independent\footnote{It is very important to emphasize that in the present formulation, we make use of the kinematic constraints (if any) to identify the truly independent fields which define the phase space in the Faddeev-Jackiw sense.} dynamical phase space fields and its coefficients $\theta_I$ are the components of the so-called canonical $1-$form (which, in general, depends on the fields)  and $H$ is the Hamiltonian. For computational purposes, it is useful to work with the Lagrangian density\footnote{Recall that, in general, the Lagrangian density depends on the fields and its space-time derivatives; in our notation, $\nabla \xi$ represent the spatial derivatives of the fields.}
\begin{equation}
L=\int \mathcal{L}(x)d^3x \quad, \quad H = \int \mathcal{H}(x)d^3x\qquad\Rightarrow\qquad 
\mathcal{L}= \theta_I \dot{\xi}^I-\mathcal{H}\;,
\end{equation}
with  $\theta_I= \theta_I (\xi;\nabla \xi)$ and $\mathcal{H}=\mathcal{H}(\xi;\nabla \xi)$. The Euler-Lagrange equations obtained from the above Lagrangian are given by \cite{andy,bufy}
\begin{equation}  
\int\omega_{IJ}(x;y)\dot\xi^J(y)d^3y=\frac{\delta H  }{\delta \xi^I(x)} \qquad , \qquad \text{with}\quad\omega_{IJ}(x;y) = \frac{\delta \theta_J(y)}{\delta \xi^I(x) }-\frac{\delta \theta_I(x)}{\delta \xi^J(y)}\;. 
\label{FJEOM}
\end{equation}
It is interesting to mention that $\omega_{IJ}$ are the components of the $2-$form obtained by taking the (functional) exterior derivative\footnote{Recall that $\delta \theta[\xi]=\displaystyle\int \delta \theta_J(y)\wedge\delta\xi^J(y)d^3y+(-1)^1\displaystyle\int\ \theta_J(y)\cancelto{0}{\delta\big(\delta\xi^J(y)\big)}\;\;\;d^3y$, with
$\delta \theta_J(y)=\displaystyle\int \dfrac{\delta\theta_J(y)}{\delta\theta_I(x)}\delta\theta_I(x)d^3x$.} of the canonical $1-$form
\begin{equation}
\theta[\xi]=\int \theta_J(y) \delta \xi^J(y) d^3y \quad\Rightarrow\quad \delta \theta = \dfrac{1}{2}\int \left(\frac{\delta \theta_J(y)}{\delta \xi^I(x) }-\frac{\delta \theta_I(x)}{\delta \xi^J(y)}\right)\delta\xi^I(x)\wedge\delta\xi^J(y)d^3xd^3y\;.
\end{equation}
Thus, we get the fundamental relation
\begin{equation}
\omega = \delta \theta \qquad ;\qquad \omega[\xi]=\dfrac{1}{2}\int\omega_{IJ}(x;y)\delta\xi^I(x)\wedge\delta\xi^J(y)d^3xd^3y\;.
\label{Fund.Rel.1}
\end{equation}
In the non-singular case, $\omega$ is the symplectic form on the canonical phase space and, since is not degenerate, possesses an inverse, with components $\omega^{IJ}$, obeying
\begin{equation}
\int \omega^{IK}(x;z)\omega_{KJ}(z;y)d^3z=\delta^I_J\delta^3(\vec{x}-\vec{y})\;,
\end{equation}
this relation can be used to express the equation of motion \eqref{FJEOM} in terms of $\omega^{IJ}$
\begin{equation}
    \dot{\xi}^I (x)=\int \omega^{IJ}(x;y)\dfrac{\delta H}{\delta \xi^J(y)}d^3y\;.
\label{FJEOM2}
\end{equation}
On the other hand, the time evolution of an arbitrary dynamical (space) functional $F$ can be obtained as follows
\begin{equation}
    \dot{F}=\int \dfrac{\delta F}{\delta\xi^I(x)}\dot{\xi}^I(x)d^3x\overset{\eqref{FJEOM2}}{=}\int\dfrac{\delta F}{\delta\xi^I(x)}\omega^{IJ}(x;y)\dfrac{\delta H}{\delta\xi^J(y)}d^3xd^3y\;.
\end{equation}
Then, considering the Hamiltonian as the generator (in the classical sense) of time evolution, according to $\dot{F}\doteq\{F;H\}$, leads us to the following definition of Poisson brackets
\begin{equation}
\lbrace F; G \rbrace \doteq \int \dfrac{\delta F}{\delta\xi^I(x)}\omega^{IJ}(x;y)\dfrac{\delta G}{\delta \xi^{J}(y)}d^3xd^3y\;,
\label{FJPB}
\end{equation}
from the above definition, one readily obtains the important result
\begin{equation}
\left\lbrace \xi^I(x) ; \xi^J(y) \right\rbrace = \omega^{IJ}(x;y)\;,
\end{equation}
which relates the elements of the matrix representation of $\omega^{-1}$ with the PB of the canonical phase space fields. This is, in fact, one of the main advantages of the FJ formulation.

In the more general case, the exterior derivative of the canonical $1-$form yields a pre-symplectic form, which implies that its matrix representation is singular. The fact that any degenerate bilinear form has a non-trivial kernel can be understood in terms of the eigenvectors corresponding to the null eigenvalue\footnote{In general, given a functional matrix $M$, we said that $v$ is an eigenvector related to the eigenvalue $r=r(x)$ if the condition $\displaystyle\int M_{IJ}(x;y)v^J(y)d^ny=r(x)v^I(x)$ is satisfied.} of the associated matrix; these eigenvectors are often called the zero modes of such a matrix. We will denote them as
\begin{equation}
v_r = \int  v_r^I(x)\dfrac{\delta}{\delta\xi^I(x)} d^3x\;,
\end{equation}
in which $r$ is a label indicating the (possible) degree degeneracy of the null eigenvalue. Making use of the equations of motion it is straightforward to check that applying $v_r$ on the Hamiltonian lead to the (dynamical) constraints
\begin{equation}
G_r = \int   v_r^I(x)\dfrac{\delta H}{\delta\xi^I(x)} d^3x \overset{!}{=}\int \mathcal{G}_r(x)d^3x=0\;,
\label{FJConst}
\end{equation}
in which we have introduced the constraint densities $\mathcal{G}_r$, since $G_r$ actually are also spatial functionals. These constraints determine a surface on the initial phase space and we must express all the quantities only on such a surface. Therefore, we write
\begin{equation}
L[\zeta,\dot{\zeta}]\overset{!}{=}L[\xi,\dot{\xi}]\Big|_{G_r=0}=\int \left(\theta_{I'}\dot{\zeta}^{I'}-\mathcal{H}\big|_{\mathcal{G}_r=0}\right)d^3x\;.
\end{equation}
In many situations, some of the fields $\xi^B$ are present in the Hamiltonian density by multiplying some of the constraints and consequently, they disappear when the constraints are imposed. Taking this into account, we have denoted the surviving fields as $\zeta^{I'}$.
Due to the dynamical character of the constraints $G_r$, we need to require their stability along the time evolution of the system; this is getting by introducing Lagrange multipliers fields 
\begin{align}
L[\zeta,\dot{\zeta}]\rightarrow L[\zeta,\dot{\zeta}]- \int\lambda^r(x) \dot{\mathcal{G}}_r(x)d^3x &\cong L[\zeta,\dot{\zeta}]+\int\dot{\lambda}^r(x) \mathcal{G}_r(x) d^3x \label{artifice}\\ 
&= \int \left(\theta_{I'}\dot{\zeta}^{I'}+ \mathcal{G}_r\dot{\lambda}^r-\mathcal{H}\big|_{\mathcal{G}_r=0}\right)d^3x\;.
\label{FJ.Lag.1}
\end{align}
The presence of the $\lambda^r$ implies an extension of the old constraint surface, with field-coordinates $\zeta^{I'}$, into a new set $M^{(1)}$, with field-coordinates $(\{\zeta^{I'}\};\{\lambda^r\})$ which we denote by $\xi^{(1)I}$. In matrix notation (which will be often used)
$\xi^{(1)}=\begin{pmatrix} \zeta^{I'} & \lambda^r \end{pmatrix}^t.$ 
Then, considering the objects
\begin{equation}
\theta^{(1)}[\xi^{(1)}]\overset{!}{=}\theta[\zeta]+\int\mathcal{G}_r(x)\delta \lambda^r(x)d^3x \qquad , \qquad \mathcal{H}^{(1)}\overset{!}{=}\mathcal{H}\big|_{\mathcal{G}_r=0}\;,
\label{1.can1f}
\end{equation}
the above Lagrangian can be written as
\begin{equation}
L[\xi^{(1)},\dot{\xi}^{(1)}]=\int \left(\theta^{(1)}_I\dot{\xi}^{(1)I}-\mathcal{H}^{(1)}\right)d^3x\equiv \int \mathcal{L}^{(1)}d^3x\;.
\label{FJ.Lag.2}
\end{equation}
Note that we have arrived at an identical expression in eq. \eqref{can.lag} and so, we are able to repeat all the above procedures. The quantities with superscript $(1)$ correspond to the so-called first iteration of the Faddeev-Jackiw algorithm. The program stops when we end up with a (non-degenerate) symplectic form $\omega^{(n)}$, since its inverse determines a Poisson structure on the (extended) phase space $M^{(n)}$.

If at some stage (iteration), say $m$, of the algorithm we obtain zero-modes of the pre-symplectic form that do not give rise to new constraints (in the sense that they result in identically null tautologies), then the system has gauge freedom. To see that, let us take an arbitrary variation of eq. \eqref{can.lag}
$$
\delta  L^{(m)} = \int\delta\xi^{(m)I}(x) \omega^{(m)}_{IJ}(x;y)\dot{\xi}^{(m)J}(y)d^3xd^3y -\delta H^{(m)}\;,
$$
with $\omega_{IJ}^{(m)}$ defined as in eq. \eqref{FJEOM}. If, in particular, we choose $\delta$ as a gauge variation, we know that the LHS of the equation becomes zero; On the other hand, it can be shown that the $m-$iterated Hamiltonian $H^{(m)}$, which takes into account all the constraints that arise among the algorithm, is gauge invariant, and therefore
\begin{equation}
\int\delta_{G}\xi^{(m)I}(x) \omega^{(m)}_{IJ}(x;y)\dot{\xi}^{(m)J}(y)d^3xd^3y=0\quad \xrightarrow{\eqref{FJEOM}}\quad\int\delta_{G}\xi^{(m)I}(x)\dfrac{\delta H^{(m)}}{\delta \xi^{(m)I}(x)}d^3x=0\;,
\label{gauge.VF}
\end{equation}
thus, we find that the vector field  $X_{G}=\displaystyle\int\delta_{G}\xi^{(m)I}(x)\dfrac{\delta}{\delta \xi^{(m)I}}d^3x$ cancels out the gauge invariant Hamiltonian $H^{(m)}$. To continue with the program, appropriate gauge conditions must be imposed by hand, replacing each identically null constraint.

Once we get a symplectic form $\omega^{(n)}$, we can determine the time evolution of all of the variables present in the (extended) phase space $M^{(n)}$ by means of eq. \eqref{FJEOM}
\begin{equation}
\int\omega^{(n)}_{IJ}(x;y)\dot\xi^{(n)J}(y)d^3y=\frac{\delta H^{(n)}  }{\delta \xi^{(n)I}(x)}
\quad\Rightarrow\quad
\dot{\xi}^{(n)I}(x)=\int \omega^{(n)IJ}(x;y)\dfrac{\delta H^{(n)}}{\delta \xi^{(n)J}(y)}d^3y\;,    
\end{equation}


The knowledge of the model phase space structure leads to a well-defined path integral measure built by means of the determinant of the matrix representation of ${\omega}^{(n)}$. The construction of the path integral is based on the Darboux theorem, seen in the discussion \cite{Toms}, for example.\\


\section{Free GS Model: A FJ Approach}
\label{sc4}

\indent In order to proceed with the FJ analysis of the GS model, we need to rewrite its Lagrangian density in the canonical form. However, because this theory has higher-order derivatives, we must be careful when defining the canonical momenta. We must notice that the complete model is a second-order theory on the fields $A_\mu$ (due to the sector $\mathcal{L}_P$), whereas it is a first-order theory on the scalar field $B$. Thus, we must employ the Ostrogadsky approach \cite{Whittaker} for the fields $A_\mu$, and the regular canonical construction for the field $B$.

\begin{itemize}
\item[$(i).$] Vector field $A_\mu$
    \item Independent fields (Ostrogradsky recipe):
    
   \begin{equation}
       A_\mu \qquad , \qquad \Gamma_\mu \doteq \dot{A}_\mu.
    \end{equation}

    \item Momentum fields (Ostrogradsky recipe)
    \begin{equation}
    \begin{cases}
        \;\;\;\;\Pi^\mu  \doteq   \dfrac{\partial \mathcal{L}}{\partial\dot{A}_\mu}-2\partial_i\left(\dfrac{\partial\mathcal{L}}{\partial(\partial_i\dot{A}_\mu)}\right)-\partial_0\left(\dfrac{\partial \mathcal{L}}{\partial\ddot{A}_\mu}\right)=F^{\mu 0}-\dfrac{1}{m^2}\left(\eta^{\mu i}\partial_i\partial_jF^{0j}-\partial_\nu\dot{F}^{\mu\nu}\right) \cr
        \;\;\;\;\Phi^\mu  \doteq\dfrac{\partial \mathcal{L}}{\partial\ddot{A}_\mu}=\dfrac{1}{m^2}\left(\eta^{\mu 0}\partial_\nu F^{0\nu}-\partial_\nu F^{\mu\nu}\right)
    \end{cases}.
    \end{equation}
    In the above $(A_\mu;\Pi^\mu)$ and $(\Gamma_\mu;\Phi^\mu)$ are canonical pairs. Besides that, note that we end up with the kinematic constraint
\begin{equation}
\Phi_0=0.
\label{kin.const}
\end{equation}
\item[$(ii).$] Scalar field $B$
    \item Conjugate momentum field
    \begin{equation}
    \pi\doteq\dfrac{\partial\mathcal{L}}{\partial \dot{B}}=\dfrac{M^2}{m_s}\left(A_0 + \dfrac{1}{m_s}\dot{B}\right)\;.
    \end{equation}
\end{itemize}
Since we have introduced new fields, we can readily obtain their corresponding gauge transformations, from eq. \eqref{gauge.tr}
\begin{equation}
 \left\lbrace
    \begin{tikzpicture}[baseline=-0.5ex]
    \matrix[
            matrix of math nodes,
             column sep=0.1ex,
            ] (m)
        {
\;\; \delta A_\mu = \partial_\mu \Lambda \quad \Rightarrow \quad \delta A_0 = \dot{\Lambda} \quad , \quad \delta A_j=\partial_j \Lambda\\
\;\; \delta \Gamma_\mu = \delta \dot{A}_\mu \quad\, \Rightarrow \quad \delta \Gamma_0 = \ddot{\Lambda} \quad , \quad \delta \Gamma_j=\partial_j \dot{\Lambda}\\
        };
    \end{tikzpicture}\right.\;;
\end{equation}
However, since $\Gamma_\mu$ are regarded as being independent of $A_\mu$, their gauge variations must be unrelated (in the sense that they should not depend on the same parameter), then
\begin{equation}
\delta \Gamma_0=\dot{\Xi} \quad , \quad \delta \Gamma_j = \partial_j\Xi \quad ; \quad \text{with} \;\;\; \Xi \overset{!}{=}\dot{\Lambda}\;.
\end{equation}
From the above, we end up with the following gauge transformations
\begin{equation}
\delta A_0 = \Xi \quad , \quad \delta \Gamma_j = \partial_j \Xi \qquad ; \qquad \delta A_j = \partial_j\Lambda \quad , \quad \delta B = -m_s \Lambda\;.
\label{gg.tr.if}
\end{equation}

According to the Ostrogadsky procedure, and taking into account eq. \eqref{kin.const}, the corresponding Hamiltonian density is given by
\begin{equation}
\mathcal{H}=\Pi^\mu\Gamma_\mu+\Phi^j\dot{\Gamma}_j+\pi\dot{B}-\mathcal{L}\;,
\label{cal.H}
\end{equation}
with
\begin{equation}
\dot{\Gamma}_\mu = m^2\Phi_\mu+\delta^0_\mu(\partial_j\partial^j A_0-\partial_j\Gamma^j)+\partial_\mu\Gamma_0+\partial^j F_{\mu j}\quad , \quad \dot{B}=m_s\left(\frac{m_s}{M^2}\pi-A_0\right)\;.
\end{equation}
After replacing these velocities in eq. \eqref{cal.H}, one gets
\begin{align}
\mathcal{H}=&\Pi^\mu\Gamma_\mu + \Phi^j\left(\dfrac{m^2}{2}\Phi_j+\partial_j\Gamma_0+\partial^kF_{jk}\right)+\left(\dfrac{m_s^2}{2M^2}\pi-m_sA_0\right)\pi+\dfrac{1}{2}(\partial_jA_0-\Gamma_j)(\partial^jA_0-\Gamma^j)+\nonumber \\
&+\dfrac{1}{4}F_{jk}F^{jk}-\dfrac{1}{2m^2}(\partial_j\Gamma^j-\partial_j\partial^j A_0)^2-\dfrac{M^2}{2}\left(A^j+\dfrac{1}{m_s}\partial^jB\right)\left(A_j+\dfrac{1}{m_s}\partial_jB\right)\;.   
\label{old.Hcal}
\end{align}
So, from eq. \eqref{cal.H} we recover the canonical form of the Lagrangian density
\begin{equation}
{\cal{L}}=\Pi^\mu \dot A_\mu + \Phi^j\dot \Gamma_j+\pi\dot B-{\cal{H}}\;.
\label{can.L0}
\end{equation}
Then, the phase space fields are identified as being
\begin{equation}
\xi=\begin{pmatrix}
A_\mu & \Pi^\mu & \Gamma_j & \Phi^j & \Gamma_0 & B & \pi
\end{pmatrix}^t.
\label{xi0}
\end{equation}
In terms of these variables, from eq. \eqref{can.L0} we construct the corresponding canonical $1-$form
\begin{equation}
\theta[\xi]=\int\Big(\Pi^\mu(y)\delta A_\mu (y) + \Phi^j (y) \delta \Gamma_j (y) +\pi (y) \delta B (y) \Big)d^3y\;.
\end{equation}
Thus, by taking the exterior derivative of $\theta$ we get the functional $2-$form
\begin{equation}
\omega[\xi]=\int \Big(\delta\Pi^\mu(x)\wedge\delta A_\mu (y) + \delta \Phi^j (x)\wedge \delta \Gamma_j (y) + \delta \pi (x)\wedge \delta B (y)\Big)\delta^3(\vec{x}-\vec{y})d^3xd^3y\;,
\end{equation}
whose functional matrix representation is
\begin{equation}
\begin{blockarray}{lccccccc}
\quad\;\boldsymbol{\wedge} & \;{\delta A_\nu (y)} &\; {\delta \Pi^\nu(y)} &\; {\delta \Gamma_k(y)} &\; {\delta \Phi^k(y)} &\; {\delta \Gamma_0(y)} &\; {\delta B(y)} &\; {\delta \pi(y)} \;\\
    \begin{block}{l(ccccccc)}
{\;\delta A_\mu (x)\quad} &\; 0 & -\delta^\mu_\nu & 0 & 0 & 0 & 0 & 0 \;\topstrut\\
{\delta \Pi^\mu(x)\quad} &\; \delta^\nu_\mu & 0 & 0 & 0 & 0 & 0 & 0 \;\\
{\,\;\delta \Gamma_j(x)\quad} &\; 0 & 0 & 0 & -\delta^j_k & 0 & 0 & 0\;\\
{\;\delta \Phi^j(x)\quad} &\; 0 & 0 & \delta^k_j & 0 & 0 & 0 & 0\;\\
{\,\;\delta \Gamma_0(x)\quad} &\; 0 & 0 & 0 & 0 & 0 & 0 & 0 \;\\
{\;\;\;\,\delta B(x)\quad} &\;0 & 0 & 0 & 0 & 0 & 0 & -1\;\\
{\;\;\;\,\delta \pi(x)\quad} &\; 0 & 0 & 0 & 0 & 0 & 1 & 0 \;\botstrut\\
    \end{block}
\end{blockarray}\quad\delta^3(\vec{x}-\vec{y}).
\end{equation}
From this expression we can easily get the column zero-mode $v^t=\begin{pmatrix} 0_\nu & 0^\nu & 0_j & 0^j & 1 & 0 & 0 \end{pmatrix}$, or in analytic notation, $v=\displaystyle\int \dfrac{\delta }{\delta \Gamma_0(x)}d^3x$. In arrangement with eq. \eqref{FJConst}, we must apply this zero-mode to the Hamiltonian to obtain the following constraint
\begin{equation}
G=\int \dfrac{\delta H}{\delta \Gamma_0(x)}d^3x= \displaystyle\int \dfrac{\delta \mathcal{H}(y)}{\delta \Gamma_0(x)}d^3xd^3y=\int \left(\Pi_0(x)-\dfrac{\partial}{\partial x^j}\Phi^j(x)\right)d^3x=0\;,
\end{equation}
from which we identify the constraint density
\begin{equation}
\mathcal{G}=\Pi_0-\partial_j\Phi^j\;.
\end{equation}
Now, we must impose this constraint on the previous system before starting with the first iteration of the FJ algorithm. To identify the remaining fields after imposing $G$ we must look at the $1-$iterated Hamiltonian density $\mathcal{H}^{(1)}=\mathcal{H}\big|_{\mathcal{G}=0}$, which, up a surface term, results
\begin{align}
\mathcal{H}^{(1)}&= \Pi^j\Gamma_j + \Phi^j\left(\dfrac{m^2}{2}\Phi_j+\partial^k F_{jk}\right)+\left(\dfrac{m_s^2}{2M^2}\pi-m_sA_0\right)\pi+\dfrac{1}{2}(\partial_jA_0-\Gamma_j)(\partial^jA_0-\Gamma^j)+\nonumber \\
& \quad\quad\quad\quad +\dfrac{1}{4}F_{jk}F^{jk}-\dfrac{1}{2m^2}(\partial_j\Gamma^j-\partial_j\partial^j A_0)^2-\dfrac{M^2}{2}\left(A^j+\dfrac{1}{m_s}\partial^jB\right)\left(A_j+\dfrac{1}{m_s}\partial_jB\right)\;.
\end{align}
Then, taking into account eq. \eqref{FJ.Lag.1} and eq. \eqref{FJ.Lag.2}, we construct the $1-$iterated canonical Lagrangian density as
\begin{equation}
\mathcal{L}^{(1)}= \Pi^\mu \dot{A}_\mu + \Phi^j\dot{\Gamma}_j+\pi \dot{B}+\mathcal{G}\dot{\lambda}-\mathcal{H}^{(1)}\;,
\end{equation}
with $\lambda=\lambda(x)$, a Lagrange multiplier field. Since $\Gamma_0$ disappears from the entire dynamics we have the surviving fields
\begin{equation}
\xi \quad\longrightarrow \quad\zeta=\xi\,\backslash \lbrace \Gamma_0\rbrace = \begin{pmatrix}
A_\mu & \Pi^\mu & \Gamma_j & \Phi^j & B & \pi
\end{pmatrix}^t.
\end{equation}
Below we present the other objects involved in the first iteration  
\begin{itemize}
\item Fields defining the functional phase space $M^{(1)}$ 
\begin{equation}
\xi^{(1)} = \begin{pmatrix} \zeta & \lambda
\end{pmatrix}^t=\begin{pmatrix}
A_\mu & \Pi^\mu & \Gamma_j & \Phi^j & B & \pi & \lambda
\end{pmatrix}^t.
\end{equation}
\item Functional canonical $1-$form:
\begin{equation}
\theta^{(1)}[\xi^{(1)}]\overset{\eqref{1.can1f}}{=}\theta[\zeta]+\int \Big(\Pi_0(y)-\partial_j^y\Phi^j(y)\Big)\delta\lambda(y) d^3y\;.
\end{equation}
\item Exterior derivative of $\theta^{(1)}$
\begin{align}
\omega^{(1)}[\xi^{(1)}]&=\omega[\zeta]+\int\Big(\delta \Pi_0(x)-\delta\Phi^j(x)\partial_j^y\Big)\delta^3(\vec{x}-\vec{y})\wedge\delta\lambda(y)d^3xd^3y\nonumber \\
&=\omega[\zeta]+\int\Big(\delta \Pi_0(x)\wedge\delta\lambda(y)+\delta\Phi^j(x)\wedge\delta\lambda(y)\partial_j^x\Big)\delta^3(\vec{x}-\vec{y})d^3xd^3y\;.
\end{align}
\item Matrix representation\footnote{Note that according to the definition of $\omega_{IJ}(x;y)$ given in eq. \eqref{FJEOM}, it turns out that $\omega_{IJ}(x;y)=-\omega_{JI}(y;x)$. We have to use this property to complete the corresponding functional matrix properly.} of $\omega^{(1)}$
\begin{equation}
\begin{pmatrix}
   0 & -\delta^\mu_\nu & 0 & 0 & 0 & 0 & 0 \cr
   \delta^\nu_\mu & 0 & 0 & 0 & 0 & 0 & \delta^0_\mu \cr
   0 & 0 & 0 & -\delta^j_k  & 0 & 0 & 0 \cr
   0 & 0 & \delta^k_j & 0  & 0 & 0 & \partial_j^x\cr
   0 & 0 & 0 & 0  & 0 & -1 & 0 \cr
   0 & 0 & 0 & 0  & 1 & 0 & 0 \cr
   0 & -\delta^0_\nu & 0 & \partial_k^x & 0 & 0 & 0 
\end{pmatrix}\delta^3(\vec{x}-\vec{y})\;.
\end{equation}
\end{itemize}

The above matrix has the following zero-mode
\begin{equation}
v^{(1)}=\begin{pmatrix}
 \delta^0_\nu \vartheta & 0^\nu & \partial_k \vartheta & 0^k & 0 & 0 & -\vartheta
\end{pmatrix}^t \equiv \int \left(\vartheta(x)\dfrac{\delta}{\delta A_0(x)}+\partial_j \vartheta(x)\dfrac{\delta}{\delta \Gamma_j(x)}-\vartheta(x)\dfrac{\delta}{\delta \lambda(x)}\right)d^3x\;,
\end{equation}
with $\vartheta=\vartheta(x)$ an arbitrary scalar function. As we know, $v^{(1)}$ give rise to a new constraint according to
\begin{equation}
G^{(1)}=\int\left(\vartheta(x)\dfrac{\delta{H}^{(1)}}{\delta A_0(x)}+\partial_j \vartheta(x)\dfrac{\delta{H}^{(1)}}{\delta \Gamma_j(x)}\right)d^3x \cong \int \vartheta(x)\Big(-m_s\pi(x)-\partial_j^x\Pi^j(x)\Big)d^3x=0\;.
\end{equation}
Due to the arbitrariness of $\vartheta$ we establish the constraint density as being:
\begin{equation}
\mathcal{G}^{(1)}=-m_s\pi-\partial_j\Pi^j\;.
\end{equation}
Then, we start the second iteration of the Faddeev-Jackiw algorithm
\begin{itemize}
\item Hamiltonian density
\begin{align}
\mathcal{H}^{(2)}&\cong \Pi^j\Gamma_j + \Phi^j\left(\dfrac{m^2}{2}\Phi_j+\partial^k F_{jk}\right)+\dfrac{m_s^2}{2M^2}\pi^2-\Pi^j\partial_jA_0+\dfrac{1}{2}(\partial_jA_0-\Gamma_j)(\partial^jA_0-\Gamma^j)+\nonumber \\
& \quad \quad\quad  +\dfrac{1}{4}F_{jk}F^{jk}-\dfrac{1}{2m^2}(\partial_j\Gamma^j-\partial_j\partial^j A_0)^2-\dfrac{M^2}{2}\left(A^j+\dfrac{1}{m_s}\partial^jB\right)\left(A_j+\dfrac{1}{m_s}\partial_jB\right)\;.
\end{align}
\item Canonical Lagrangian density
\begin{equation}
\mathcal{L}^{(2)}=\Pi^\mu \dot{A}_\mu + \Phi^j\dot{\Gamma}_j+\pi \dot{B}+\mathcal{G}\dot{\lambda}+\mathcal{G}^{(1)}\dot{\lambda}^{(1)}-\mathcal{H}^{(2)}\;.
\end{equation}
We should note that in this iteration none of the fields were suppressed, so $\zeta^{(1)}=\xi^{(1)}$.
\item Fields defining the functional phase space $M^{(2)}$ 
\begin{equation}
\xi^{(2)}=\begin{pmatrix}
 \xi^{(1)} & \lambda^{(1)}
\end{pmatrix}^t.
\end{equation}
\item Functional canonical $1-$form
\begin{equation}
\theta^{(2)}[\xi^{(2)}] = \theta^{(1)}[\xi^{(1)}]-\int \Big(m_s\pi+\partial_j^y\Pi^j\Big)\delta\lambda^{(1)}(y)d^3y\;.
\end{equation}
\item Exterior derivative of $\theta^{(2)}$
\begin{equation}
\omega^{(2)}[\xi^{(2)}]=\omega^{(1)}[\xi^{(1)}]-\int\Big(m_s\delta \pi(x)-\delta\Pi^j(x)\partial_j^x\Big)\delta^3(\vec{x}-\vec{y})\wedge\delta\lambda^{(1)}(y)d^3x d^3y \;.
\end{equation}
\item Matrix representation of $\omega^{(2)}$
\begin{equation}
\begin{pmatrix}
0 & -\delta^\mu_\nu & 0 & 0 & 0 & 0 & 0 & 0 \cr
\delta^\nu_\mu & 0 & 0 & 0 & 0 & 0 & \delta^0_\mu & \delta^\ell_\mu\partial^x_\ell\cr
0 & 0 & 0 & -\delta^j_k  & 0 & 0 & 0 & 0\cr
0 & 0 & \delta^k_j & 0  & 0 & 0 & \partial_j^x & 0 \cr
0 & 0 & 0 & 0  & 0 & -1 & 0 & 0 \cr
0 & 0 & 0 & 0  & 1 & 0 & 0 & -m_s\cr
0 & -\delta^0_\nu & 0 & \partial_k^x & 0 & 0 & 0 & 0 \cr
0 & \delta^\ell_\nu \partial^x_\ell & 0 & 0 & 0 & m_s & 0 & 0 
\end{pmatrix}\delta^3(\vec{x}-\vec{y})\;.
\end{equation}
\end{itemize}
The above matrix possesses the following column zero-modes
\begin{equation}
\begin{cases}
\;\;v_1^{(2)}=
\begin{pmatrix}
\delta^0_\nu\vartheta & 0^\nu & \partial_k\vartheta & 0^k & 0 & 0 &  -\vartheta & 0
\end{pmatrix}^t\cr
\;\;v_2^{(2)}=
\begin{pmatrix}
\delta_\nu^\ell\partial_\ell\sigma & 0^\nu & 0_k & 0^k & -m_s\sigma  & 0 & 0 & -\sigma
\end{pmatrix}^t
\end{cases},
\end{equation}
whose analytic expressions are given by
\begin{equation}
\begin{cases}
\;\; v_1^{(2)} =\displaystyle\int\left(\vartheta(x)\dfrac{\delta}{\delta A_0(x)}+\partial_j\vartheta(x)\dfrac{\delta}{\delta\Gamma_j(x)}-\vartheta(x)\dfrac{\delta}{\delta\lambda(x)}\right)d^3x \cr
\;\; v_2^{(2)} =\displaystyle\int\left(\partial_j\sigma(x)\dfrac{\delta}{\delta A_j(x)}-m_s\sigma(x)\dfrac{\delta}{\delta B(x)}-\sigma(x)\dfrac{\delta}{\delta\lambda^{(1)}(x)} \right)d^3x
\end{cases}.
\end{equation}
The constraints generated by these zero-modes should be
\begin{equation} 
G^{(2)}_1=v_1^{(2)}H^{(2)}=0 \qquad , \qquad G^{(2)}_2=v_2^{(2)}H^{(2)}=0\;.
\end{equation}
However, after computing explicitly these expressions, tautologies of type $0=0$ are obtained and, therefore, do not represent new constraints. Note that $v_{1,2}^{(2)}H^{(2)}=0$  have the same form of the second equation in eq. \eqref{gauge.VF}, with $v_1^{(2)}$ and $v_2^{(2)}$ both playing the role of $X_G$. This is, clearly, a necessary condition for the equations of motion to be invariant under the following local transformations 
\begin{equation}
\begin{cases}
\;\;\delta A_0 = \vartheta \qquad \;\;\; , \qquad \delta \Gamma_j=\partial_j\vartheta \qquad \;\;\,  , \qquad \delta\lambda = -\vartheta \cr
\;\;\delta A_j = \partial_j\sigma \qquad , \qquad \delta B=-m_s\sigma \qquad , \qquad \delta\lambda^{(1)} = \sigma
\end{cases}\;;
\label{inf.gg.tr}
\end{equation}
which are compatible with the gauge transformations presented in eq. \eqref{gg.tr.if}, with $\Xi\leftrightarrow \vartheta$ and $\Lambda\leftrightarrow \sigma$. Since we have obtained two tautologically null constraints, we must impose two additional conditions\footnote{Properly, we use an (achievable) gauge condition and a non-dynamical restriction which arises from the consistency of the former.} to replace them. For this task, we use the Coulomb-Podolsky-Stueckelberg gauge condition, which is consistently attainable\footnote{To see this, it is sufficient to replace $\Lambda=\left[\left(1+\dfrac{\square}{m^2}\right)\nabla^2-M^2\right]^{-1}\left\lbrace \left(1+\dfrac{\square}{m^2}\right)\nabla\cdot\vec{A}-\dfrac{M^2}{m_s}B\right\rbrace$ in the gauge transformations.}, which is the more general (non-covariant) condition in our an Ostrogadskian framework
\begin{equation}  
\mathcal{G}^{(2)}_1=\left(1+\frac{\square}{m^2}\right)\partial_k A^k-\frac{M^2}{m_s}B=0 \;,
\end{equation}
which in turn fix the dynamics of $A_0$ by means of its equation of motion, \begin{equation}\mathcal{G}^{(2)}_2=\left[M^2-\left(1+\dfrac{\square}{m^2}\right)\nabla^2\right]A_0=0\;.\end{equation}

Hence, we end up with these two extra conditions that will lead to an (invertible) symplectic form. By repeating the steps of the FJ algorithm, the following symplectic form is obtained
\begin{align}
\omega^{(3)}[\xi^{(3)}]&=\omega^{(2)}[\xi^{(2)}]-\int \left[\delta A_j(x)\left(1+\dfrac{\square_x}{m^2}\right)\partial^j_x+\dfrac{M^2}{m_s}\delta B(x)\right]\delta^3(\vec{x}-\vec{y})\wedge\delta \lambda_1^{(2)}(y)d^3x d^3y+\nonumber\\
& \qquad \qquad \qquad \qquad \qquad  +\int \left[M^2-\left(1+\dfrac{\square_x}{m^2}\right)\nabla^2_x\right]\delta^3(\vec{x}-\vec{y})\delta A_0(x)\wedge \delta \lambda_2^{(2)}(y)d^3x d^3y\;, 
\end{align}
with $\xi^{(3)}=\begin{pmatrix} \xi^{(2)} & \lambda^{(2)}_1 & \lambda^{(2)}_2 \end{pmatrix}^t$. The matrix representation of $\omega^{(3)}$ is 
\begin{equation}
\left(\begin{array}{cccccccccc}
 0 & -\delta^\mu_\nu & 0 & 0 & 0 & 0 & 0 & 0 & -\delta^\mu_\ell\mathbf{K}\partial^\ell & \delta^\mu_0\mathbf{G}\cr
 \delta^\nu_\mu & 0 & 0 & 0 & 0 & 0 & \delta^0_\mu & \delta^\ell_\mu\partial_\ell & 0 & 0 \cr
 0 & 0 & 0 & -\delta^j_k  & 0 & 0 & 0 & 0 & 0 & 0 \cr
 0 & 0 & \delta^k_j & 0  & 0 & 0 & \partial_j & 0 & 0 & 0 \cr
 0 & 0 & 0 & 0  & 0 & -1 & 0 & 0 & -M^2/m_s & 0 \cr
 0 & 0 & 0 & 0  & 1 & 0 & 0 & -m_s & 0 & 0 \cr
 0 & -\delta^0_\nu & 0 & \partial_k & 0 & 0 & 0 & 0 & 0 & 0  \cr
 0 & \delta^\ell_\nu \partial_\ell & 0 & 0 & 0 & m_s & 0 & 0 & 0 & 0 \cr  
-\delta^\nu_\ell\mathbf{K}\partial^\ell & 0 & 0 & 0 & M^2/m_s & 0 & 0 & 0 & 0 & 0 \cr
-\delta^\nu_0\mathbf{G} & 0 & 0 & 0 & 0 & 0 & 0 & 0 & 0 & 0
\end{array}\right)\delta^3(\vec{x}-\vec{y})\;,
\end{equation}
in which all the differential operators just involve derivatives with respect to $x$ and we have called $\mathbf{K}\overset{!}{=}1+\dfrac{\square}{m^2}$ and $\mathbf{G}\overset{!}{=}M^2-\mathbf{K}\nabla^2$. The inverse of the functional above matrix is given by\footnote{It is important to mention that the inverse object of a (functional) symplectic form is called a (functional) Poisson bivector field and is written in the basis $\left\lbrace\frac{\delta}{\delta \xi^I(x)}\wedge\frac{\delta}{\delta \xi^J(y)}\right\rbrace$; unlike the symplectic form, which is commonly written in the basis $\left\lbrace\delta \xi^I(x)\wedge\delta\xi^J(y)\right\rbrace$.}

\begin{equation}
\left(\mkern-1mu
\begin{tikzpicture}[baseline=-0.5ex]
\matrix[
  matrix of math nodes,
  column sep=1ex,
  ](m)
  {
0 & \mathbf{M}_\mu{}^\nu & 0 & 0 & 0 & \frac{\delta^\ell_\mu M^2 \partial_\ell}{m_s\mathbf{G}} & 0 & 0 & -\frac{\delta^\ell_\mu\partial_\ell}{\mathbf{G}} & -\frac{\delta^0_\mu}{\mathbf{G}}\\
-\mathbf{M}_\nu{}^\mu & 0 & -\delta^\mu_0\partial_k & 0 & \frac{\delta^\mu_\ell m_s\mathbf{K}\partial^\ell}{\mathbf{G}} & 0 & -\delta^\mu_0 & \frac{\delta^\mu_\ell\mathbf{K}\partial^\ell}{\mathbf{G}} & 0 & 0 \\
0 & -\delta^\nu_0\partial_j & 0 & \delta^k_j & 0 & 0 & 0 & 0 & 0 & -\frac{\partial_j}{\mathbf{G}} \\
0 & 0 & -\delta^j_k & 0 & 0 & 0 & 0 & 0 & 0 & 0 \\
0 & \frac{\delta^\nu_\ell m_s\mathbf{K}\partial^\ell}{\mathbf{G}} & 0 & 0 & 0 & -\frac{\mathbf{K}\nabla^2}{\mathbf{G}} & 0 & 0 & \frac{m_s}{\mathbf{G}} & 0 \\
\frac{\delta^\ell_\nu M^2 \partial_\ell}{m_s\mathbf{G}} & 0 & 0 & 0 & \frac{\mathbf{K}\nabla^2}{\mathbf{G}} & 0 & 0 & \frac{M^2}{m_s\mathbf{G}} & 0 & 0 \\    
0 & \delta^\nu_0 & 0 & 0 & 0 & 0 & 0 & 0 & 0 & \frac{1}{\mathbf{G}} \\ 
0 & \frac{\delta^\nu_\ell\mathbf{K}\partial^\ell}{\mathbf{G}} & 0 & 0 & 0 & -\frac{M^2}{m_s\mathbf{G}} & 0 & 0 & \frac{1}{\mathbf{G}} & 0 \\
-\frac{\delta^\ell_\nu\partial_\ell}{\mathbf{G}} & 0 & 0 & 0 & -\frac{m_s}{\mathbf{G}} & 0 & 0 & -\frac{1}{\mathbf{G}} & 0 & 0 \\  \noalign{\vskip 0.3ex}
\frac{\delta^0_\nu}{\mathbf{G}} & 0 & -\frac{\partial_k}{\mathbf{G}} & 0 & 0 & 0 & -\frac{1}{\mathbf{G}} & 0 & 0 & 0\\
};
\draw[dashed]
  ([xshift=2.5ex]m-1-8.north east) -- ([xshift=2.5ex]m-10-8.south east);
\draw[dashed]
  ([yshift=-1ex]m-8-1.south west) -- ([yshift=-1ex]m-8-10.south east);
\end{tikzpicture}\mkern-1mu
\right)\delta^3(\vec{x}-\vec{y})\;,
\end{equation}
with $\mathbf{M}_\mu{}^\nu\overset{!}{=}\delta^j_\mu\delta^\nu_k\left(\delta^k_j  - \dfrac{\mathbf{K}\partial_j\partial^k}{\mathbf{G}}\right)$. Note that the rows and columns corresponding to the $\lambda-$variables (Lagrange multiplier fields) were separated by dashed lines. In this way, the correct Poisson brackets of the surviving dynamical phase space fields will be given by the entries in the first block. Thus, the non-vanishing equal-time Poisson brackets are
\begin{align}
\lbrace A_j(x);\Pi^k(y)\rbrace = \left[\mathbf{M}_j{}^k\right]_x\delta^3(\vec{x}-\vec{y}) \quad &,\quad \lbrace A_j(x);\pi(y)\rbrace = \dfrac{M^2}{m_s}\dfrac{\partial_j^x}{\mathbf{G}_x}\delta^3(\vec{x}-\vec{y}) \quad , \nonumber \\ 
\lbrace \Pi^0(x);\Gamma_k(y)\rbrace = -\partial_k^x\delta^3(\vec{x}-\vec{y}) \quad & ,\quad \lbrace \Pi^j(x);B(y)\rbrace = m_s \dfrac{\mathbf{K}_x}{\mathbf{G}_x}\partial^j_x\delta^3(\vec{x}-\vec{y}) \quad ,\label{F.PB}\\
\lbrace \Gamma_j(x);\Phi^k(y)\rbrace = \delta^k_j\delta^3(\vec{x}-\vec{y}) \quad &,\quad \lbrace B(x);\pi(y)\rbrace = -\dfrac{\mathbf{K}_x}{\mathbf{G}_x}\nabla^2_x\delta^3(\vec{x}-\vec{y})\quad. \nonumber
\end{align}

The determinant of $\omega^{(3)}$ turns out to be proportional to $\det\left( \mathbf{G}^4\right)$, which is indeed the same value obtained by the DB algorithm in the approach of \cite{m2y}. It can be used to build the path integral measure. {As well as in the present formalism, there is recent interest in a wide set of alternative phase space studies. For example, one may cite \cite{refref}, also in the context of higher derivative theories, in which the extended Chern-Simons theory coupled to scalar matter is analyzed, revealing an interesting structure for the Poisson brackets.}


\section{Path Integral Quantization in the FJ framework}
\label{sc5}

Since the last iteration of the Faddeev-Jackiw algorithm provides a symplectic structure for the (extended) phase space $M^{(n)}$, the corresponding Lagrangian density $\mathcal{L}^{(n)}$ becomes non-singular and therefore the path integral approach can be implemented.
In general, the generating functional of correlation functions in the absence of sources will be given by an expression of the form
\begin{equation}
Z = N\int \exp\left\lbrace i \int \mathcal{L}^{(n)}(x)d^4x\right\rbrace\mathscr{D}\mu  \equiv N \int \exp\left\lbrace iS^{(n)}[\xi^{(n)}]\right\rbrace\mathscr{D}\mu \qquad , \qquad \mathscr{D}\mu=\mathscr{D}\mu[\xi^{(n)}]\;;
\label{ansatz}
\end{equation}
in which $\mathscr{D}\mu$ is a functional measure of integration to be determined and $N$ is just a normalization factor. At this point, it is worth recalling that if $\Omega$ is the standard\footnote{That is, the one whose matrix representation has the form $\begin{pmatrix}
0 & -\mathbb{I} \cr \mathbb{I} & 0 \end{pmatrix}\delta^3(\vec{x}-\vec{y}).$} functional symplectic form, then it is written in the so-called Darboux coordinates, say $\Xi$. Then, if we perform a change in the phase space $M^{(n)}$ between the $\Xi-$coordinates and the $\xi^{(n)}$ ones, we will have
\begin{equation}
\omega^{(n)}_{IJ}(x;y)=\int \dfrac{\delta \Xi^M(x')}{\delta\xi^{(n)I}(x)}\Omega_{MN}(x';y')\dfrac{\delta \Xi^N(y')}{\delta\xi^{(n)J}(y)}d^3x'd^3y'\;,
\end{equation}
since this equation corresponds to the components of the multiplication of functional matrices, we can write the above relation as
\begin{equation}
\omega^{(n)} = \left(\dfrac{\delta\Xi}{\delta\xi^{(n)}}\right)^t \Omega \left(\dfrac{\delta\Xi}{\delta\xi^{(n)}}\right)\;,
\end{equation}
from this equation, we recognize the presence of the Jacobian functional matrix corresponding to the transformation $\Xi\to\xi^{(n)}$. Then, taking the determinant in both sides of the above equation yields
\begin{equation}
\det\left(\omega^{(n)}\right)=\det\left(\dfrac{\delta\Xi}{\delta\xi^{(n)}}\right)\det(\Omega)\det\left(\dfrac{\delta\Xi}{\delta\xi^{(n)}}\right)=\det{}^2\left(\dfrac{\delta\Xi}{\delta\xi^{(n)}}\right)\;.
\label{aux1}
\end{equation}
On the other hand, we know that in the Darboux coordinates, the generating functional adopts the simple form
\begin{equation}
Z = N\int \exp\Big\lbrace i S[\Xi]\Big\rbrace\mathscr{D}[\Xi] = N\int \exp\left\lbrace i \int \mathcal{L}\big(\Xi(x);\partial_\mu\Xi(x)\big)d^4x\right\rbrace\mathscr{D}[\Xi]\;;
\label{aux2}
\end{equation}
then, the change $\Xi\rightarrow\xi^{(n)}$ implies the measure transformation
\begin{equation}
\mathscr{D}[\Xi]= \left|\det\left(\dfrac{\delta\Xi}{\delta\xi^{(n)}}\right)\right|\mathscr{D}[\xi^{(n)}]\overset{\eqref{aux1}}{=} \left| \det\left(\omega^{(n)}\right)\right|^{1/2}\mathscr{D}[\xi^{(n)}]\;,
\end{equation}
Thus, replacing this into \eqref{aux2} yields the expression for the generating functional with integrand in terms of the $\xi^{(n)}-$fields
\begin{equation}
Z =  \int \exp\left\lbrace i \int \mathcal{L}^{(n)}\big(\xi^{(n)}(x);\partial_\mu\xi^{(n)}(x)\big)d^4x\right\rbrace\left| \det\left(\omega^{(n)}\right)\right|^{1/2}\mathscr{D}[\xi^{(n)}]\;.
\label{gener.Z}
\end{equation}
Hence, by comparing with \eqref{ansatz} we obtain the desired integration measure
\begin{equation}
\mathscr{D}\mu[\xi^{(n)}]=\left| \det\left(\omega^{(n)}\right)\right|^{1/2}\mathscr{D}[\xi^{(n)}]\;.
\end{equation}
Therefore, we see that the Faddeev-Jackiw approach provides the correct measure for the functional integral $Z$. On the other hand, we know that the $\xi^{(n)}-$fields have the following structure
\begin{equation}
\xi^{(n)}=\begin{pmatrix}
\zeta & \lambda & \lambda^{(1)} & \cdots & \lambda^{(n)}
\end{pmatrix}^t\;,
\end{equation}
in which we have denoted by $\zeta^{M}$ the \textit{dynamical fields}; i. e. those remaining after having imposed all the dynamical constraints\footnote{Of course, this is not the case for subsidiary gauge conditions, if any.} which arose along the algorithm in the Hamiltonian density. However, these fields are not necessarily independent because there could have been non-independent fields in the \textit{canonical sector}\footnote{With this we are referring to the term which contains the velocities in the canonical Lagrangian}, since this part is not evaluated on the constraints surface (unlike the Hamiltonian density). With this notation, we can write the $n-$iterated canonical Lagrangian density as
\begin{equation}
\mathcal{L}^{(n)}=\theta_M\dot{\zeta}^M+\dot{\lambda}\mathcal{G}+\dot{\lambda}^{(1)}\mathcal{G}^{(1)}+\cdots+\dot{\lambda}^{(n)}\mathcal{G}^{(n)}-\mathcal{H}^{(n)}\;.
\end{equation}
Recall that the original proposal in the Faddeev-Jackiw approach was to impose the time-consistency of the constraints; however, we made the artifice in \eqref{artifice} just for convenience's sake. Then, we are free to reverse it
\begin{equation}
\mathcal{L}^{(n)}=\theta_M\dot{\zeta}^M-\mathcal{H}^{(n)}-\left(\lambda\dot{\mathcal{G}}+\lambda^{(1)}\dot{\mathcal{G}}^{(1)}+\cdots+\lambda^{(n)}\dot{\mathcal{G}}^{(n)}\right)\;.
\end{equation}
Replacing this expression for $\mathcal{L}^{(n)}$ in \eqref{gener.Z} we get
\begin{eqnarray}
&&Z = N\int \exp\left\lbrace i\int\left(\theta_M\dot{\zeta}^M-\mathcal{H}^{(n)}\right)d^4x\right\rbrace \exp\left\lbrace -i\int\left(\lambda\dot{\mathcal{G}}+\lambda^{(1)}\dot{\mathcal{G}}^{(1)}+\cdots+\lambda^{(n)}\dot{\mathcal{G}}^{(n)}\right)d^4x\right\rbrace\times\cr\cr
&& \hspace{7.5cm}\times\left| \det\left(\omega^{(n)}\right)\right|^{1/2}\mathscr{D}[\lambda,\lambda^{(1)},\cdots,\lambda^{(n)}]\mathscr{D}[\zeta]\;,
\end{eqnarray}
with $\mathscr{D}[\zeta]\overset{!}{=}\displaystyle\prod_{M}\mathscr{D}[\zeta^M]$. It is easy to check that $\det\left(\omega^{(n)}\right)$ does not depend on the $\lambda-$variables and therefore we can isolate the functional integral on these variables as follows
\begin{align}
Z &= N\int \exp\left\lbrace i\int\left(\theta_M\dot{\zeta}^M-\mathcal{H}^{(n)}\right)d^4x\right\rbrace \left| \det\left(\omega^{(n)}\right)\right|^{1/2}\left(\prod_{k=0}^n\int \exp\left\lbrace -i\int\lambda^{(k)}\dot{\mathcal{G}}^{(k)}d^4x\right\rbrace \mathscr{D}[\lambda^{(k)}]\right)\mathscr{D}[\zeta] \nonumber\\
&=N\int \exp\left\lbrace i\int\left(\theta_M\dot{\zeta}^M-\mathcal{H}^{(n)}\right)d^4x\right\rbrace \left| \det\left(\omega^{(n)}\right)\right|^{1/2}\delta[\,\dot{\mathcal{G}}\,]\,\delta[\dot{\mathcal{G}}^{(1)}]\cdots\delta[\dot{\mathcal{G}}^{(n)}]\mathscr{D}[\zeta]\;.
\end{align}
Note that each functional delta can be putted in the form $\delta[\dot{\mathcal{G}}^{(k)}]=\Big(\,\det(\frac{\partial}{\partial t})\,\Big)^ {-1}\delta [\mathcal{G}^{(k)}]$, and since $\det(\frac{\partial}{\partial t})$ does not depend on the dynamical fields $\zeta$, it can be absorbed by the normalization factor\footnote{This absorption is possible since $\det(\frac{\partial}{\partial t})$ can be expressed as a functional integral over (real) Grassmann-valued fields $\mathcal{C}$ and $\bar{\mathcal{C}}$ as: $\displaystyle\int \exp\left\lbrace -\int \bar{\mathcal{C}}(x)\frac{\partial}{\partial t}\delta^4(x-x'){\mathcal{C}(x')}d^4x\,d^4x' \right\rbrace\mathscr{D}[{\mathcal{C}},\bar{\mathcal{C}}]$, and there is no coupling between these auxiliary fields (commonly known as \textit{ghosts}) and the dynamical fields after replacing into the expression for $Z$.}; i. e., $N\rightarrow N \det(\frac{\partial}{\partial t})$. Therefore
\begin{equation}
Z = N \int \exp\left\lbrace i\int\left(\theta_M\dot{\zeta}^M-\mathcal{H}^{(n)}\right)d^4x\right\rbrace \left| \det\left(\omega^{(n)}\right)\right|^{1/2}\delta[\,\mathcal{G}\,]\,\delta[\mathcal{G}^{(1)}]\cdots\delta[\mathcal{G}^{(n)}]\mathscr{D}[\zeta]\;.
\label{GBAppr}
\end{equation}
In our case, according \eqref{GBAppr}, the generating functional will given by the expression\footnote{Since from the definition of $\xi^{(3)}$ one identifies the dynamical fields are given by $\zeta=\begin{pmatrix} A_\mu & \Pi^\mu & \Gamma_j & \Phi^j & B & \pi\end{pmatrix}^t$, we have explicitly:
$\mathscr{D}[\zeta]\equiv\left(\displaystyle\prod_{\mu=0}^3\mathscr{D}[A_\mu]\right)\left(\displaystyle\prod_{\mu=0}^3\mathscr{D}[\Pi^\mu]\right)\left(\displaystyle\prod_{j=1}^3\mathscr{D}[\Gamma_j]\right)\left(\displaystyle\prod_{j=1}^3\mathscr{D}[\Phi^j]\right)\mathscr{D}[B]\mathscr{D}[\pi]$.}
\begin{equation}
Z = N\int\exp\left\lbrace i\int \mathcal{L}^{(3)}_{dyn}d^4x\right\rbrace\left|\det\left(\omega^{(3)}\right)\right|^{1/2}\delta[\,\mathcal{G}\,]\delta[\mathcal{G}^{(1)}]\delta[\mathcal{G}^{(2)}_1]\delta[\mathcal{G}^{(2)}_2]\mathscr{D}[A_\mu , \Pi^\mu , \Gamma_j, \Phi^j, B , \pi]\;,
\end{equation}
with
\begin{align}
&\mathcal{L}^{(3)}_{dyn} = \Pi_0\dot{A}_0 + \Pi^j\dot{A}_j + \Phi^j\dot{\Gamma}_j+\pi\dot{B}-\mathcal{H}^{(3)}\;, \label{L3}\\
\mathcal{G}=\Pi_0-\partial_j \Phi^j \quad, \quad &\mathcal{G}^{(1)}=-m_s\pi-\partial_j\Pi^j \quad, \quad \mathcal{G}^{(2)}_1=\mathbf{K}\nabla\cdot\vec{A}-\dfrac{M^2}{m_s}B \quad, \quad \mathcal{G}^{(2)}_2=\mathbf{G}A_0
\;. \nonumber
\end{align}
Taking into account that the $3-$iterated Hamiltonian density is equal to
\begin{eqnarray}
&\mathcal{H}^{(3)}=\Pi^j\Gamma_j + \dfrac{m^2}{2}\Phi^j\Phi_j+\Phi^j\partial^k F_{jk}+\dfrac{m_s^2}{2M^2}\pi^2-\Pi^j\partial_j A_0+\dfrac{1}{2}(\Gamma_j-\partial_jA_0)(\Gamma^j-\partial^jA_0)+\cr\cr
&\qquad \qquad -\dfrac{1}{2m^2}\Big(\partial_j(\Gamma^j-\partial^jA_0)\Big)^2+\dfrac{1}{4}F_{jk}F^{jk}-\dfrac{M^2}{2}\left(A_j+\partial_j\dfrac{B}{m_s}\right)\left(A^j+\partial^j\dfrac{B}{m_s}\right)\;, \label{H3}
\end{eqnarray}
we get the following expression for the generating functional
\begin{eqnarray}
&Z=N\int\exp\bigg\lbrace i\int[\Pi_0\dot{A}_0 + \Phi^j\left(\dot{\Gamma}_j-\partial^k F_{jk}\right)- \dfrac{m^2}{2}\Phi^j\Phi_j+\Pi^j(\dot{A}_j-\Gamma_j+\partial_j A_0) +\cr\cr
&\hspace{2cm} -\dfrac{1}{2}(\Gamma_j-\partial_jA_0)(\Gamma^j-\partial^jA_0)+ +\dfrac{1}{2m^2}\Big(\partial_j(\Gamma^j-\partial^jA_0)\Big)^2 +\pi\dot{B}-\dfrac{m_s^2}{2M^2}\pi^2+\cr\cr
&\hspace{2cm} -\dfrac{1}{4}F_{jk}F^{jk}+\dfrac{M^2}{2}\left(A_j+\partial_j\dfrac{B}{m_s}\right)\left(A^j+\partial^j\dfrac{B}{m_s}\right)]d^4x\bigg\rbrace\left|\det(\mathbf{G}^4)\right|^{1/2}\delta[\Pi_0-\partial_j \Phi^j]\times \nonumber \cr\cr
&\hspace{7.5cm}\times\delta[\mathcal{G}^{(1)}]\delta[\mathcal{G}^{(2)}_1]\delta[\mathcal{G}^{(2)}_2]\mathscr{D}[A_\mu , \Pi^\mu, \Gamma_j, \Phi^j, B , \pi]\\.
\end{eqnarray}
As a first step, it is convenient to perform the following integral
\begin{align}
\int\exp\left\lbrace i\int \Pi_0\dot{A}_0d^4x\right\rbrace\delta[\Pi_0-\partial_j \Phi^j]\mathscr{D}[\Pi_0]=\exp\left\lbrace i\int \left(\partial_j\Phi^j\right)\dot{A}_0d^4x\right\rbrace\cong
\exp\left\lbrace -i\int \Phi^j\partial_j\dot{A}_0d^4x\right\rbrace\;
\end{align}
With this, the expression for $Z$ reduces to
\begin{eqnarray}
&&Z=N\int\exp\bigg\lbrace i\int[\Phi^j\left(\dot{\Gamma}_j-\partial_j\dot{A}_0-\partial^k F_{jk}\right)- \dfrac{m^2}{2}\Phi^j\Phi_j+\Pi^j(\dot{A}_j-\Gamma_j+\partial_j A_0)+\cr\cr
&&\hspace{2cm}- \dfrac{1}{2}(\Gamma_j-\partial_jA_0)(\Gamma^j-\partial^jA_0)+\dfrac{1}{2m^2}\Big(\partial_j(\Gamma^j-\partial^jA_0)\Big)^2+\pi\dot{B}-\dfrac{m_s^2}{2M^2}\pi^2-\dfrac{1}{4}F_{jk}F^{jk}+\cr\cr
&&\hspace{2cm}+\dfrac{M^2}{2}\left(A_j+\partial_j\dfrac{B}{m_s}\right)\left(A^j+\partial^j\dfrac{B}{m_s}\right)]d^4x\bigg\rbrace\left|\det(\mathbf{G}^4)\right|^{1/2}\delta[\mathcal{G}^{(1)}]\,\delta[\mathcal{G}^{(2)}_1]\,\delta[\mathcal{G}^{(2)}_2]\times\cr\cr
&&\hspace{11cm}\times\mathscr{D}[A_\mu , \Pi^j, \Gamma_j, \Phi^j, B , \pi]\;.\nonumber \\
\end{eqnarray}
Note that since $\det(\mathbf{G}^4)$ does not depend on the fields in the integration measure, could be absorbed by the normalization factor; however, we will keep that factor. On the other hand, it results convenient to call $\Theta_j\overset{!}{=}\Gamma_j-\partial_j A_0$. It is easy to check that the corresponding Jacobian is equal to $1$ and then
\begin{eqnarray}
&&Z=N \det{}^2(\mathbf{G})\int\exp\bigg\lbrace i\int\bigg[\Phi^j\left(\dot{\Theta}_j-\partial^k F_{jk}\right)- \dfrac{m^2}{2}\Phi^j\Phi_j-\Pi^j(\Theta_j-\dot{A}_j) - \dfrac{1}{2}\Theta_j\Theta^j+\dfrac{1}{2m^2}\Big(\partial_j\Theta^j\Big)^2+\cr\cr
&&\hspace{4cm}+\pi\dot{B}-\dfrac{m_s^2}{2M^2}\pi^2-\dfrac{1}{4}F_{jk}F^{jk}+\dfrac{M^2}{2}\left(A_j+\partial_j\dfrac{B}{m_s}\right)\left(A^j+\partial^j\dfrac{B}{m_s}\right)\bigg]d^4x\bigg\rbrace\times\cr\cr
&&\hspace{5.6cm}\times\delta[-m_s\pi-\partial_j\Pi^j]\delta[\mathcal{G}^{(2)}_1]\delta[\mathbf{G}(A_0)]\mathscr{D}[A_\mu , \Pi^j, \Theta_j, \Phi^j, B , \pi]\;.
\end{eqnarray}
The following step consists in performing the following integral
\begin{align}
I&=\int\exp\left\lbrace-i\int\Pi^j(\Theta_j-\dot{A}_j)d^4x\right\rbrace\delta[-m_s\pi-\partial_j\Pi^j]\mathscr{D}[\Pi^j] \nonumber \\
&\sim \int\exp\left\lbrace-i\int\Pi^j(\Theta_j-\dot{A}_j)d^4x\right\rbrace\left(\int\exp\left\lbrace i\int(m_s\pi+\partial_j\Pi^j)\sigma d^4x\right\rbrace\mathscr{D}[\sigma]\right)\mathscr{D}[\Pi^j] \nonumber \\
&\cong \int\exp\left\lbrace-i\int\Pi^j(\Theta_j-\dot{A}_j+\partial_j\sigma)d^4x\right\rbrace \exp\left\lbrace i\int m_s\pi\sigma d^4x\right\rbrace\mathscr{D}[\sigma,\Pi^j] \nonumber \\
&\sim\int \delta[\Theta_j-\dot{A}_j+\partial_j\sigma]\exp\left\lbrace i\int m_s\pi\sigma d^4x\right\rbrace\mathscr{D}[\sigma]\;.
\label{Auxint1}
\end{align}
On the other hand, consider the following integral
\begin{equation}
\int \delta[\mathbf{G}(A_0)]\mathscr{D}[A_0]=\int\Big(det(\mathbf{G})\Big)^{-1}\delta[A_0]\mathscr{D}[A_0]=\Big(det(\mathbf{G})\Big)^{-1}\;
\label{GW1}
\end{equation}
With these considerations, the expression for the generating functional becomes
\begin{eqnarray}
&&Z=N\det(\mathbf{G})\int\exp\left\lbrace i\int\left[\Phi^j\left(\dot{\Theta}_j-\partial^k F_{jk}\right)-\dfrac{m^2}{2}\Phi^j\Phi_j-\dfrac{1}{2}\Theta_j\Theta^j+\dfrac{1}{2m^2}\Big(\partial_j\Theta^j\Big)^2+\right.\right.\cr\cr
&&\hspace{1.5cm}+\pi(\dot{B}+m_s\sigma)-\dfrac{m_s^2}{2M^2}\pi^2-\dfrac{1}{4}F_{jk}F^{jk}\left.\left.+\dfrac{M^2}{2}\left(A_j+\partial_j\dfrac{B}{m_s}\right)\left(A^j+\partial^j\dfrac{B}{m_s}\right)\right]d^4x\right\rbrace \times\cr\cr
&&\hspace{6.5cm}\times\delta[\Theta_j-\dot{A}_j+\partial_j\sigma]\delta[\mathcal{G}^{(2)}_1]\mathscr{D}[\sigma, A_j, \Theta_j, \Phi^j, B , \pi]\; \nonumber \\
\end{eqnarray}
Now, we perform the following integral
\begin{align}
I'&=\int\exp\left\lbrace i\int\left[\Phi^j\dot{\Theta}_j-\dfrac{1}{2}\Theta_j\Theta^j+\dfrac{1}{2m^2}\Big(\partial_j\Theta^j\Big)^2\right]d^4x\right\rbrace\delta[\Theta_j-\dot{A}_j+\partial_j\sigma]\mathscr{D}[\Theta_j]\nonumber \\
&=\exp\left\lbrace i\int\left[\Phi^j(\ddot{A}_j-\partial_j\dot{\sigma})-\dfrac{1}{2}(\dot{A}_j-\partial_j\sigma)(\dot{A}^j-\partial^j\sigma)+\dfrac{1}{2m^2}\Big(\partial_j(\dot{A}^j-\partial^j\sigma)\Big)^2\right]d^4x\right\rbrace\;.
\end{align}
The relabeling $\sigma\rightarrow A_0$ allows us to write $(\dot{A}_j-\partial_j \sigma)\rightarrow (\dot{A}_j-\partial_j A_0) \equiv F_{0j}$, and therefore:
\begin{equation}
I'=\exp\left\lbrace i\int\left[\Phi^j\dot{F}_{0j}-\dfrac{1}{2}F_{0j}F^{0j}+\dfrac{1}{2m^2}\Big(\partial_jF^{0j}\Big)^2\right]d^4x\right\rbrace\;
\end{equation}
On the other hand, the integration over the field $\pi$ yields
\begin{equation}
\int\exp\left\lbrace i\int\left[\pi(\dot{B}+m_s\sigma)-\dfrac{m_s^2}{2M^2}\pi^2\right]d^4x\right\rbrace\mathscr{D}[\pi]\sim \exp\left\lbrace i\int \dfrac{M^2}{2m_s^2}(\dot{B}+m_sA_0)^2d^4x\right\rbrace\;.
\end{equation}
With this, we arrive at the following form for $Z$
\begin{eqnarray}
&&Z= N\det(\mathbf{G})\int\exp\left\lbrace i\int\left[\Pi^j(\dot{F}_{0j}-\partial^kF_{jk})-\dfrac{m^2}{2}\Phi^j\Phi_j+\dfrac{1}{2m^2}\Big(\partial_jF^{0j}\Big)^2-\dfrac{1}{2}F_{0j}F^{0j}-\dfrac{1}{4}F_{jk}F^{jk}+\right.\right.\cr\cr
&&\hspace{1.75cm}\left.\left.+\dfrac{M^2}{2m_s^2}(m_sA_0+\dot{B})^2+\dfrac{M^2}{2}\left(A_j+\partial_j\dfrac{B}{m_s}\right)\left(A^j+\partial^j\dfrac{B}{m_s}\right)\right]d^4x\right\rbrace 
\delta[\mathcal{G}^{(2)}_1]\mathscr{D}[A_\mu, \Phi^j, B ]\cr\cr
&&=N\det(\mathbf{G})\int\exp\left\lbrace i\int\left[\Pi^j(\dot{F}_{0j}-\partial^kF_{jk})-\dfrac{m^2}{2}\Phi^j\Phi_j+\dfrac{1}{2m^2}\Big(\partial_jF^{0j}\Big)^2-\dfrac{1}{4}F_{\mu\nu}F^{\mu\nu}\right.\right.+\cr\cr
&&\hspace{4cm}\left.\left.+\dfrac{M^2}{2}\left(A_\mu+\partial_\mu\dfrac{B}{m_s}\right)\left(A^\mu+\partial^\mu\dfrac{B}{m_s}\right)\right]d^4x\right\rbrace 
\delta[\mathcal{G}^{(2)}_1]\mathscr{D}[A_\mu, \Phi^j, B ]
\;.
\end{eqnarray}
By performing the integral over the fields $\Phi^j$, we get
\begin{align}
    &\int\exp\left\lbrace i\int\left[\Pi^j(\dot{F}_{0j}-\partial^kF_{jk})-\dfrac{m^2}{2}\Phi^j\Phi_j\right]d^4x\right\rbrace \mathscr{D}[\Phi^j]\cr
    &\hspace{2.5cm}\sim \exp\left\lbrace i\int \dfrac{1}{2m^2}\left(\dot{F}_{0j}-\partial^k F_{jk}\right)\left(\dot{F}^{0j}-\partial_k F^{jk} \right)d^4x\right\rbrace\;
\end{align}
Therefore
\begin{eqnarray}
\small
&&Z=N\det(\mathbf{G})\int\exp\left\lbrace i\int\left[\dfrac{1}{2m^2}\left(\dot{F}_{0j}+\partial^k F_{kj}\right)\left(\dot{F}^{0j}+\partial_k F^{kj} \right)+\dfrac{1}{2m^2}\partial_j F^{j0}\partial^k F_{k0}+\right.\right.\cr\cr
&&\hspace{1.4cm}\left.\left.-\dfrac{1}{4}F_{\mu\nu}F^{\mu\nu}+\dfrac{M^2}{2}\left(A_\mu+\partial_\mu\dfrac{B}{m_s}\right)\left(A^\mu+\partial^\mu\dfrac{B}{m_s}\right)\right]d^4x\right\rbrace 
\delta[\mathcal{G}^{(2)}_1]\mathscr{D}[A_\mu, B ]\;
\end{eqnarray}
thus, we arrive at the following expression
\begin{multline}
    Z=N\det(\mathbf{G})\int\exp\left\lbrace i\int\left[-\dfrac{1}{4}F_{\mu\nu}F^{\mu\nu}+\frac{1}{2m^2}\partial^\mu F_{\rho\mu}\partial_\nu F^{\rho\nu}\right.\right.+\nonumber \\
    \left.\left.-\dfrac{M^2}{2}\left(A_\mu+\partial_\mu\dfrac{B}{m_s}\right)\left(A^\mu+\partial^\mu\dfrac{B}{m_s}\right)\right]d^4x\right\rbrace 
\delta[\mathcal{G}^{(2)}_1]\mathscr{D}[A_\mu, B ]\;
\end{multline}
which in turn can be written in a more suggestive form
\begin{equation}
    Z=N\det(\mathbf{G})\int\exp\Big\lbrace i\,S[A_\mu,B]\Big\rbrace \delta\left[\left(1+\dfrac{\square}{m^2}\right)\partial^jA_j-\dfrac{M^2}{m_s}B\right]\mathscr{D}[A_\mu, B ]\;.
    \label{noncov.Z.free}
\end{equation}
Although this amplitude is correct, it is not manifestly covariant. Then, in order to turn it into a
covariant form, we should take into account the following:

\begin{itemize}
    \item[$(i).$]We can write the involved gauge condition in a more convenient way:
    \begin{eqnarray}
    F(A_{\mu};B)=(1+\frac{\Box}{m^{2}})\partial^{j}A_{j}-\frac{M^{2}}{m_{s}}B=(1+\frac{\Box}{m^{2}})\partial^{\mu}A_{\mu}-\frac{M^{2}}{m_{s}}B-(1+\frac{\Box}{m^{2}})\dot{A}_{0}=0.
    \end{eqnarray}
    \item[$(ii).$]Considering this authentic gauge condition, we can make use the Faddeev-Popov (FP) procedure and inserting the following identity object into any path integral with functional dependence on the fields $A_{\mu}$ and $B$:
    \begin{eqnarray}
    1=\det\left(\frac{\delta}{\delta\Lambda}F(A_{\mu}^{\Lambda};B^{\Lambda})\right)|_{\Lambda=0}\int \delta[F(A_{\mu}^{\Lambda};B^{\Lambda})]\mathscr{D}[\Lambda],
    \end{eqnarray}
    with $A_{\mu}$ and $B$ being the transformed gauge fields according to (\ref{gauge.tr}).
     \item[$(iii).$]Using the fact that the gauge invariance of the FP-determinant, the action and the functional measure involving the gauge fields allows us to \textit{reverse} the gauge transformed field, we end
up with an expression of the form:
\begin{eqnarray}
&&\int ... e^{iS[A_{\mu},B]}\mathscr{D}[A_{\mu},B]=\left(\int \mathscr{D}[\Lambda] \right)\int ...e^{iS[A_{\mu},B]}\det[M^{2}-(1+\frac{\Box}{m^{2}})\nabla^{2}]\times\cr\cr
&&\times\delta\left[ (1+\frac{\Box}{m^{2}})\partial^{j}A_{j}-\frac{M^{2}}{m_{s}}B\right]\mathscr{D}[A_{\mu},B]
    \end{eqnarray}
    in which the braced first factor on the RSH is a infinite constant corresponding to the so-called gauge group volume\footnote{In fact, a better definition for the generating functional describing systems with gauge invariance is the usual ones but divided by the associated gauge group volume. However, although ill-defined, for our purposes we will allow this infinite constant to be absorbed by the normalization factor.
}. Also note that the argument in the determinant correspond to the differential operator {\bf G}, previously defined, and due to its independence from the fields, it can be separated from the functional integral.
\end{itemize}

With this, the final form for the generating functional reads:
\begin{equation}
Z=N\int\exp\left\{iS[A_{\mu},B]\right\}\mathscr{D}[A_{\mu},B]
\end{equation}
The above expression reveals two very important aspects, proper of an Abelian gauge theory:
First, the extra factors coming from the initial functional measure end out being dropped due the
covariant-like fixation of the (non-covariant) gauge. Second, the resulting expression coincides
with the \textit{naive} gauge redundant ones, up to the gauge volume group and the factor $\det({\bf G})$.


\section{FJ Theory Involving Grassmann-valued Fields}
\label{sc6}

In this section, we present an extension to the Faddeev-Jackiw technique for the case in which we are dealing with Grassmann variables \cite{Manav}. As in the previous case, the construction will be done in the context of classical fields. As we saw, the $\xi-$variables correspond to the ordinary phase space fields (involving, in most cases, both configuration and momentum fields). In the present case we also will consider anticommutative (Grassmann-valued) fields, which will be called $\eta-$variables. Due to the (degree $2$) nilpotency of the Grassmann variables, the terms in the Lagrangian density containing velocities of this kind of fields will be always linear in them, and consequently, the conjugate canonical momenta of these variables turn out to be proportional to the anticommutative configuration fields and so, the $\eta-$variables only contain configuration fields. In order to distinguish the $\xi-$variables from the $\eta-$variables we may adopt the standard terminology in Grassmann calculus: The commutative objects are said to have \textit{even} Grassmann-parity ($|\xi^I|=0$), while the anticommutative objects have \textit{odd} parity ($|\eta^\alpha|=1$); alternately, we can use the physics terminology, in which we refer the $\xi-$variables as \textit{bosonic} and the $\eta-$variables, as \textit{fermionic}.\\

In this approach we define the unified variables $z^A\doteq (\lbrace \xi^I\rbrace ; \lbrace \eta^\alpha\rbrace)$, or in matrix notation: $z^{st} = \begin{pmatrix} \xi & \eta\end{pmatrix}$. The starting point is, as we know, the Lagrangian written in its canonical form\footnote{If we compare this expression with \eqref{can.lag}, we note that the velocities were put on the left of the components of the canonical $1-$form, this is because we are employing the left derivative convention.}
\begin{equation}
L[z,\dot{z}] = \int \dot{z}^A(x) \theta_A(x) d^3x - H[z]\;,
\end{equation}
this Lagrangian always have even Grassmann-parity ($|L|=0$) and as a consequence
\begin{equation}
|\theta_A| = |A| \qquad , \qquad |H|=0 \qquad ; \qquad \text{with $\;\;\;\;|A|\overset{!}{=}|z^A|\;$.}
\end{equation}  
The corresponding equations of motion result
\begin{equation}
\int \omega_{AB}(x;y)\dot{z}^B(y)d^3y = \dfrac{\delta_{_L}H}{\delta z^A(x)}\quad\; , \;\ \text{with}\quad \omega_{AB}(x;y)=(-1)^{|B|}\left(\dfrac{\delta_{_L}\theta_B(y)}{\delta z^A(x)}-(-1)^{|A||B|}\dfrac{\delta_{_L}\theta_A(x)}{\delta z^B(y)}\right)\;.	
\label{FJGrEOM}
\end{equation} 
From the explicit form of $\omega_{AB}$ we get the following properties:
\begin{equation}
\omega_{AB}(x;y)=(-1)^{(|A|+1)(|B|+1)}\omega_{BA}(y;x) \qquad , \qquad |\omega_{AB}(x;y)| = |A|+|B|\;.
\label{Prop.Omega}
\end{equation}
In our  convention, the functional exterior derivative acts on the left, i.e.: $\delta = \displaystyle\int \delta z^B(x)\dfrac{\delta_{_L}}{\delta z(x)}d^3x$, and has odd parity ($|\delta|=1$), the later corresponds to the Bernstein-Leites convention
\begin{equation}
\begin{cases}
\;\;\alpha \wedge \beta = (-1)^{|\alpha||\beta|}\beta\wedge\alpha \quad \Rightarrow \quad \delta z^A\wedge \delta z^B = (-1)^{(|A|+1)(|B|+1)}\delta z^B\wedge \delta z^A \cr
\;\; \qquad \qquad \qquad \;\; \delta (\alpha\wedge\beta) = \delta \alpha\wedge\beta + (-1)^{|\alpha|}\alpha\wedge\delta\beta
\end{cases}.
\end{equation}
Under this convention, computing the exterior derivative of the canonical $1-$form $\theta[z]= \displaystyle\int \delta z^B(y) \theta_B(y)d^3y$ yields
\begin{eqnarray}
\delta \theta &=&\int(-1)^{|B|+1}\delta z^B(y)\wedge \delta z^A(x) \dfrac{\delta_{_L}\theta_B(y)}{\delta z^A(x)}d^3xd^3y\cr\cr
&=&\dfrac{1}{2}\int(-1)^{|B|+1}\delta z^B(y)\wedge \delta z^A(x) \left(\dfrac{\delta_{_L}\theta_B(y)}{\delta z^A(x)}-(-1)^{|A||B|}\dfrac{\delta_{_L}\theta_A(x)}{\delta z^B(y)}\right)d^3xd^3y\;,
\end{eqnarray}
and hence $\omega=-\delta \theta $. This relation is in apparent contradiction with \eqref{Fund.Rel.1}, due to the negative sign; nevertheless, we must take into account two details:
\begin{itemize}
\item In the present convention, the ordering of the contractions of the component indices with those of the base elements is contrary to the ordinary case. As a consequence, in the above notation, $\delta z^A(x)$ and $\delta z^B(y)$ correspond to the $A-$th row and the $B-$th column of the (super)matrix representation of $\omega$, respectively. 
\item Since we have made the labeling $A\overset{!}{=}(I,\alpha)$ and $B\overset{!}{=}(J,\beta)$, we have\footnote{We have omitted the graded antisymmetrization and made use of the condensed De Witt notation just to avoid a cumbersome expression.}
\begin{equation}
\omega = \delta z^J\wedge\delta z^I \dfrac{\delta_{_L}\theta_J}{\delta z^I}-\delta z\,^\beta\wedge\delta z^I \dfrac{\delta_{_L}\theta_\beta}{\delta z^I}+\delta z^J\wedge\delta z^\alpha \dfrac{\delta_{_L}\theta_J}{\delta z^\alpha}-\delta z\,^\beta\wedge\delta z^\alpha \dfrac{\delta_{_L}\theta_\beta}{\delta z^\alpha}\;.
\label{DW}
\end{equation}
Ignoring the odd variables $z^\alpha$ and $z\,^\beta$ in \eqref{DW}, we stay only with the first term, and hence, according to the previous item, we recover the result of the ordinary case. 
\end{itemize}  
In the non-singular case $\omega$ is an even supersymplectic form and the components $\omega^{AB}(x;y)$ of its inverse are such that
\begin{equation}
\omega^{AB}(x;y) = -(-1)^{|A||B|}\omega^{BA}(y;x) \qquad , \qquad |\omega^{AB}(x;y)|=|A|+|B|
\end{equation}
Note that, by comparing with \eqref{Prop.Omega}, the $\omega^{AB}$ are antisupercommutative, whereas the $\omega_{AB}$ are skew-supercommutative. Like in the ordinary case, the $\omega^{AB}$ allows us to introduce the super Poisson Brackets of two dynamical superfunctions $F$ and $G$ as being
\begin{equation}
\lbrace F; G \rbrace \doteq \int \dfrac{\delta_{_R}F}{\delta z^A(x)}\omega^{AB}(x;y)\dfrac{\delta_{_L}G}{\delta z^B(y)}d^3xd^3y\;
\end{equation}
which in turn implies
\begin{equation}
\lbrace z^A(x) ; z^B(y) \rbrace = \omega^{AB}(x;y)\;
\end{equation}
In the singular case, the $2-$form $\omega$ has zero-modes $v_r=\displaystyle\int v^A_r(x)\dfrac{\delta_{_L}}{\delta z^A(x)}d^3x$; which yields the constraints when applied to the Hamiltonian as in \eqref{FJConst}. A direct consequence of this is that the constraints (and also the constraint densities) always have even Grassmann-parity and therefore also the corresponding Lagrange multipliers $\lambda^r$ that allow introducing the consistency of those in the dynamics. The rest of the algorithm is essentially identical to the ordinary case but with the caveat that, since $|\lambda^r|=0$ the must be inserted in the phase superspace $M^{(1)}$ at the end (or beginning) of the even sector and not simply at the end of the old supercoordinates, i. e.
\begin{equation}
[z^{(1)}]^{st}= \begin{pmatrix}
\zeta & \lambda & \eta
\end{pmatrix}\;.
\end{equation}
Besides that, it is important to take into account that since this case carries a $\mathbb{Z}_2-$graded structure, we are going to deal with (even) supermatrices and therefore we have to use the well-established rules for them. With these new issues, we will proceed to determine the canonical structure of the interacting case of the Podolsky-Stueckelberg model with massive Dirac fermions.

\section{GS Electrodynamics (interacting case)}
\label{sc7}

The Lagrangian density for the interacting case can be obtained from the minimal coupling prescription
\begin{equation}
\mathcal{L} = \mathcal{L}_{F} + \dfrac{i}{2}\left\lbrace \bar{\psi}_\alpha[\gamma^\mu]^\alpha{}_\beta D_\mu\psi^\beta- D_\mu\bar{\psi}_\alpha[\gamma^\mu]^\alpha{}_\beta\psi^\beta\right\rbrace-m_e\bar{\psi}_\alpha\psi^\alpha\;,
\end{equation}
in which $\mathcal{L}_F$ is the Lagrangian density presented in \eqref{FPS} and the other terms correspond to Dirac Lagrangian density with the gauge covariant derivatives
\begin{equation}
D_\mu \psi^\alpha \doteq (\partial_\mu + ie A_\mu)\psi^\alpha \qquad , \qquad 
D_\mu \psi^\dagger_\alpha \doteq (\partial_\mu - ie A_\mu)\psi^\dagger_\alpha\;;
\end{equation} 
being $e$ the electric charge of the fermion. Also, $m_e$ is the mass of the fermion. Thus, we arrive at the following Lagrangian density
\begin{equation}
\mathcal{L} = \mathcal{L}_{F} + \mathcal{L}_{D} -e A_\mu\bar{\psi}_\alpha[\gamma^\mu]^\alpha{}_\beta \psi^\beta\;,
\label{IPS}
\end{equation}
with $\mathcal{L}_{D}$ being the free Dirac Lagrangian density. In our approach, each component of the Dirac field is a complex Grassmann-valued map, i. e.: $\psi^\alpha:\mathbb{R}^{1,3}\to \mathbb{C}_a$ and so $|\psi^\alpha|=1$. The Lagrangian \eqref{IPS} is gauge invariant under the transformations in \eqref{gauge.tr} and the additional finite local $U(1)-$transformations
\begin{equation}
\psi^\alpha \rightarrow \psi'^\alpha = e^{-ie\Lambda}\psi^\alpha  \qquad , \qquad \psi^\dagger_\alpha \rightarrow \psi'^\dagger_\alpha = e^{ie\Lambda}\psi^\dagger_\alpha\;.
\label{fer.gg.tr}
\end{equation}
The corresponding odd canonical momentum fields are given by
\begin{equation}
\pi^\dagger_\alpha = \dfrac{\partial_{_L}\mathcal{L}}{\partial\dot{\psi}^\alpha}=-\dfrac{i}{2}\psi_\alpha^\dagger \qquad , \qquad
\pi^\alpha = \dfrac{\partial_{_L}\mathcal{L}}{\partial\dot{\psi}^\dagger_\alpha}=-\dfrac{i}{2}\psi^\alpha\;.
\end{equation}
Due to the dependence between the fermion fields and its momenta, we establish the phase superspace fields as being $z^{st}=\begin{pmatrix}
\xi & \psi^\alpha & \psi^\dagger_\alpha \end{pmatrix}$, with $\xi$ defined in \eqref{xi0}\footnote{In general, we will take advantage of the results obtained in the free case at each iteration of the algorithm.}. The corresponding Hamiltonian density is given by
\begin{equation}
\mathcal{H} = \mathcal{H}_{F}-\dfrac{i}{2}\left(\bar{\psi}_\alpha[\gamma^k]^\alpha{}_\beta\partial_k\psi^\beta-\partial_k\bar{\psi}_\alpha[\gamma^k]^\alpha{}_\beta\psi^\beta\right)+m_e\bar{\psi}_\alpha\psi^\alpha+eA_\mu\bar{\psi}_\alpha[\gamma^\mu]^\alpha{}_\beta\psi^\beta\;.
\end{equation}
Then, the canonical Lagrangian density results
\begin{equation}
\mathcal{L} = \mathcal{L}_{F}-\dfrac{i}{2}\dot{\psi}^\alpha\psi^\dagger_\alpha-\dfrac{i}{2}\dot{\psi}^\dagger_\alpha\psi^\alpha \equiv \mathcal{L}_{F} + \mathcal{L}_{_{ODD}}\;.
\end{equation}
From this, we construct the canonical $1-$form as being
\begin{equation}
\theta = \theta_F - \dfrac{i}{2}\int \left(\delta\psi^\alpha(y)\psi^\dagger_\alpha(y)+\delta\psi^\dagger_\alpha(y)\psi^\alpha(y)\right)d^3y
\equiv \theta_F + \theta_{_{ODD}}\;.\end{equation}
By using \eqref{DW} we obtain the $2-$form:
\begin{align}
\omega &= \omega_F + \dfrac{i}{2}\int\left(\delta\psi^\alpha(y)\wedge\delta\psi^\dagger_\alpha(x)+\delta\psi^\dagger_\alpha(y)\wedge\delta\psi^\alpha(x)\right)\delta^3(\vec{x}-\vec{y})d^3xd^3y\\
&= \omega_F + \int i\delta^\alpha_\beta\,\delta\psi^\beta(y)\wedge\delta\psi^\dagger_\alpha(x)\delta^3(\vec{x}-\vec{y})d^3xd^3y\equiv\omega_F+\omega_{_{ODD}}\;,
\end{align}
whose supermatrix representation is\footnote{As usual, the lines are separating the blocks which define the (even) supermatrix.}
\begin{equation}
\left(\mkern-4mu
\begin{tikzpicture}[baseline=-.65ex]
\matrix[
  matrix of math nodes,
  column sep=1ex,
] (m)
{
 &  & \quad\; &  &  &  \\
 & \omega_F &  &  & 0 &  \\
\quad\; &  &  &  &  &\quad\; \\
 &  &  & 0 &  & i\delta_\alpha^\beta \\
 & 0 &  &  &  &  \\
 & &\quad\; & i\delta_\alpha^\beta &  & 0 \\
};
\draw[-]
 ([xshift=1ex]m-1-3.north east) -- ([xshift=1ex]m-6-3.south east);
\draw[-]
 ([yshift=-0.4ex]m-3-1.south west) -- ([yshift=-0.4ex]m-3-6.south east);
\end{tikzpicture}\mkern-5mu
\right)\delta^3(\vec{x}-\vec{y})\;,
\end{equation}
which is singular with zero-mode $v=\displaystyle\int \dfrac{\delta}{\delta\Gamma_0(x)}d^3x\equiv v_{F}$. The constraint obtained by making $G=vH$ coincides with the got in the free case since there is no coupling between $\Gamma_0$ and the fermionic fields. Thus, we can start with the first iteration of the Faddeev-Jackiw algorithm:
\begin{itemize}
\item Hamiltonian density
\begin{equation}
\mathcal{H}^{(1)}=\mathcal{H}^{(1)}_F - \dfrac{i}{2}\left(\bar{\psi}_\alpha[\gamma^k]^\alpha{}_\beta\partial_k\psi^\beta-\partial_k\bar{\psi}_\alpha[\gamma^k]^\alpha{}_\beta\psi^\beta\right)+m_e\bar{\psi}_\alpha\psi^\alpha+eA_\mu\bar{\psi}_\alpha[\gamma^\mu]^\alpha{}_\beta\psi^\beta\;.
\end{equation}
\item Fields defining the phase superspace $M^{(1)}:$
\begin{equation}
[z^{(1)}]^{st} = \begin{pmatrix}\xi^{(1)} & \psi^\alpha & \psi^\dagger_\alpha\end{pmatrix}\;.
\end{equation}
\item Canonical Lagrangian density
\begin{equation}
\mathcal{L}^{(1)}=\mathcal{L}^{(1)}_F+\mathcal{L}_{_{ODD}}\;.
\end{equation}
\item Canonical $1-$form 
\begin{equation}
\theta^{(1)}=\theta^{(1)}_F + \theta_{_{ODD}}\;.
\end{equation}
\item Negative of the exterior derivative of $\theta^{(1)}$
\begin{equation}
\omega^{(1)}=\omega^{(1)}_F + \omega_{_{ODD}} \equiv \left(
\mkern-4mu
\begin{tikzpicture}[baseline=-.65ex]
\matrix[
  matrix of math nodes,
  column sep=1ex,
] (m)
{
 &  & \quad\; &  &  &  \\
 & \omega_F^{(1)} &  &  & 0 &  \\
\quad\; &  &  &  &  &\quad\; \\
 &  &  & 0 &  & i\delta_\alpha^\beta \\
 & 0 &  &  &  &  \\
 & &\quad\; & i\delta_\alpha^\beta &  & 0 \\
};
\draw[-]
 ([xshift=1ex]m-1-3.north east) -- ([xshift=1ex]m-6-3.south east);
\draw[-]
 ([yshift=-0.4ex]m-3-1.south west) -- ([yshift=-0.4ex]m-3-6.south east);
\end{tikzpicture}\mkern-5mu
\right)\delta^3(\vec{x}-\vec{y})\;.
\end{equation}
\end{itemize}
The supermatrix representation of $\omega^{(1)}$ turns out to be singular. Then, we have
\begin{itemize}
\item Zero-mode of $\omega^{(1)}$
\begin{equation}
v^{(1)}=\int \left(\vartheta(x)\dfrac{\delta}{\delta A_0(x)}+\partial_j \vartheta(x)\dfrac{\delta}{\delta \Gamma_j(x)}-\vartheta(x)\dfrac{\delta}{\delta \lambda(x)}\right)d^3x\equiv v^{(1)}_F\;.
\end{equation}
\item Generated constraint
\begin{eqnarray}
G^{(1)}&=&v^{(1)}H^{(1)}=G^{(1)}_F + \int \vartheta(x)\dfrac{\delta}{\delta A_0(x)}\left\lbrace eA_0(y)\psi^\dagger_\alpha(y)\psi^\alpha(y)\right\rbrace d^3xd^3y\cr\cr
&=&G^{(1)}_F+\int \vartheta(x)e\psi^\dagger_\alpha(x)\psi^\alpha(x)d^3x\;.
\end{eqnarray}
\item Constraint density
\begin{equation}
\mathcal{G}^{(1)}=\mathcal{G}^{(1)}_F + e \psi^\dagger_\alpha \psi^\alpha\;.
\end{equation}
\end{itemize}
With this, we start the second iteration of the Faddeev-Jackiw algorithm
\begin{itemize}
\item Hamiltonian density:
\begin{equation}
\mathcal{H}^{(2)}=\mathcal{H}^{(2)}_F - \dfrac{i}{2}\left(\bar{\psi}_\alpha[\gamma^k]^\alpha{}_\beta\partial_k\psi^\beta-\partial_k\bar{\psi}_\alpha[\gamma^k]^\alpha{}_\beta\psi^\beta\right) + m_e\bar{\psi}_\alpha\psi^\alpha + e A_k \bar{\psi}_\alpha [\gamma^k]^\alpha{}_\beta\psi^\beta\;.
\end{equation}
\item Fields defining the phase superspace $M^{(2)}$
\begin{equation}
[z^{(2)}]^{st} = \begin{pmatrix}\xi^{(2)} & \psi^\alpha & \psi^\dagger_\alpha\end{pmatrix}\;.
\end{equation}
\item Canonical Lagrangian density
\begin{equation}
\mathcal{L}^{(2)}=\mathcal{L}^{(2)}_F+\dot{\lambda}^{(1)}e\psi^\dagger_\alpha \psi^\alpha+\mathcal{L}_{_{ODD}}\;.
\end{equation}
\item Canonical $1-$form
\begin{equation}
\theta^{(2)}=\theta^{(2)}_F + \int \delta\lambda^{(1)}(y)\left(e\psi^\dagger_\alpha(y) \psi^\alpha(y)\right)d^3y+\theta_{_{ODD}}\;.
\end{equation}
\item Negative of the exterior derivative of $\theta^{(2)}$
\begin{equation}
\omega^{(2)}=\omega^{(2)}_F +\int\delta\lambda^{(1)}(y)\wedge\left[\delta\psi^\dagger_\alpha(x)e\psi^\alpha(y)-\delta\psi^\alpha(x)e\psi^\dagger_\alpha(y)\right]\delta^3(\vec{x}-\vec{y})d^3xd^3y + \omega_{_{ODD}} \;.
\end{equation}
\item Supermatrix representation of $\omega^{(2)}$
\begin{equation}
\left(\mkern-5mu
\begin{tikzpicture}[baseline=-.65ex]
\matrix[
  matrix of math nodes,
  column sep=1ex,
] (m)
{
   &                                    &    \qquad\quad                        &   0                         & 0                 \\         
   &   \;\;\;\;\;\;\;\;\omega^{(2)}_F   &                               &  \vdots                     & \vdots                \\
  \quad\; &                                    &                               & -e\psi^\dagger_\beta(x)     & e\psi^\beta(x)         \\
 0 &              \cdots                &   -e\psi^\dagger_\alpha(y)    &         0                   & i\delta_\alpha^\beta   \\
 0 &              \cdots                &   e\psi^\alpha(y)             &  i\delta^\alpha_\beta       &  0                     \\
 };
 \draw[-]
 ([xshift=0.5ex]m-1-3.north east) -- ([xshift=0.5ex]m-5-3.south east);
\draw[-]
 ([yshift=-1.5ex]m-3-1.south west) -- ([yshift=-1ex]m-3-5.south east);
\end{tikzpicture}\mkern-5mu
 \right)\delta^3(\vec{x}-\vec{y})\;.
\end{equation}
\end{itemize}
It is easy to check that the supermatrix representation of $\omega^{(2)}$ is singular, with zero-modes
\begin{equation}
v_1^{(2)}=v^{(2)}_{1,\,F} \qquad , \qquad v_2^{(2)}=v^{(2)}_{2,\,F} - \int \sigma(x)\left(ie\psi^\alpha(x)\dfrac{\delta_{_L}}{\delta\psi^\alpha(x)}- ie\psi^\dagger_\alpha(x)\dfrac{\delta_{_L}}{\delta\psi^\dagger_\alpha(x)}\right)d^3x\;.
\end{equation}
Nevertheless, these zero-modes do not generate new constraints, but tautologically null tautologies, and therefore two gauge conditions must be implemented. On the other hand, this fact suggests to us that the infinitesimal variations for the fermionic fields, 
\begin{equation}
\delta \psi^\alpha=-ie\sigma\psi^\alpha \qquad \text{and}  \qquad \delta \psi^\dagger_\alpha=ie\sigma\psi^\dagger_\alpha\;,
\label{small.fer.tr}
\end{equation}
along with the presented in \eqref{inf.gg.tr}, left invariant the Lagrangian density \eqref{IPS}. In fact, we shall notice that \eqref{small.fer.tr} corresponds to the infinitesimal versions of \eqref{fer.gg.tr}, with $\Lambda\rightarrow\sigma$. As in the free case, we choose the Coulomb-Podolsky-Stueckelberg gauge condition and the underlying equation of motion for $A_0$
\begin{equation}
\begin{cases}
\;\;\mathcal{G}^{(2)}_1=\left(1+\dfrac{\square}{m^2}\right)\partial_k A^k - \dfrac{M^2}{m_s}B\equiv\mathcal{G}^{(2)}_{1,\,F}\cr  \noalign{\vskip 0.65ex}
\;\;\mathcal{G}^{(2)}_2=\left[M^2-\left(1+\dfrac{\square}{m^2}\right)\nabla^2\right]A_0 - e\psi^\dagger_\alpha \psi^\alpha\equiv\mathcal{G}^{(2)}_{2,\,F}-e\psi^\dagger_\alpha \psi^\alpha
\end{cases}\;.
\end{equation}
By repeating the steps of the Faddeev-Jackiw algorithm we get the $2-$form
\begin{equation}
\omega^{(3)}=\omega^{(3)}_F-\int\delta\lambda_2^{(2)}(y)\wedge\left[\delta\psi^\dagger_\alpha(x)e\psi^\alpha(y)-\delta\psi^\alpha(x)e\psi^\dagger_\alpha(y)\right]\delta^3(\vec{x}-\vec{y})d^3xd^3y+\omega_{_{ODD}}\;.
\end{equation}
This yields the following supermatrix representation
\begin{equation}
\left(
\mkern-5mu
\begin{tikzpicture}[baseline=-.65ex]
\matrix[
  matrix of math nodes,
  column sep=1ex,
] (m)
{
   &        &                &         &      \qquad\quad\;\;\;\;    &   0                      & 0 \\
   &        &                &         &           &  \vdots                  & \vdots       \\
   &        & \;\;\;\;\;\;\omega^{(3)}_F &      &  & -e\psi^\dagger_\beta(x)  & e\psi^\beta(x)   \\
   &        &                &         &           &   0                      & 0 \\
  \quad\; &        &                &         &           &  e\psi^\dagger_\beta(x)  & -e\psi^\beta(x)   \\
 0 & \cdots & -e\psi^\dagger_\alpha(y) & 0 & e\psi^\dagger_\alpha(y) & 0  & i\delta_\alpha^\beta  \\
 0 & \cdots &  e\psi^\alpha(y)         & 0 & -e\psi^\alpha(y)        &  i\delta^\alpha_\beta &  0 \\
};
\draw[-]
  ([xshift=0.5ex]m-1-5.north east) -- ([xshift=0.5ex]m-7-5.south east);
\draw[-]
  ([yshift=-1.5ex]m-5-1.south west) -- ([yshift=-1ex]m-5-7.south east);
\end{tikzpicture}\mkern-5mu
\right)\delta^3(\vec{x}-\vec{y})\;.
\end{equation}
Following the rules of inverting an even supermatrix, we obtain:
\begin{equation}
\left(\mkern-4mu
\begin{tikzpicture}[baseline=-.65ex]
\matrix[
  matrix of math nodes,
  column sep=1ex,
] (m)
{
 &  & \quad\; &  &  &  \\
 & \left[\omega_F^{(3)}\right]^{-1} &  &  & R &  \\
\quad\; &  &  &  &  &\quad\; \\
 &  &  & 0 &  & i\delta_\alpha^\beta \\
 & S &  &  &  &  \\
 & &\quad\; & i\delta_\alpha^\beta &  & 0 \\
};
\draw[-]
 ([xshift=1ex]m-1-3.north east) -- ([xshift=1ex]m-6-3.south east);
\draw[-]
 ([yshift=-0.4ex]m-3-1.south west) -- ([yshift=-0.4ex]m-3-6.south east);
\end{tikzpicture}\mkern-5mu
\right)\delta^3(\vec{x}-\vec{y})\;,
\end{equation}
In which $R$ and $S$ are non-square blocks with odd entries given by:
\begin{equation}
R = \begin{pmatrix}
ie\psi^\beta(y)\frac{\delta^0_\mu}{\mathbf{G}} & -ie\psi^\dagger_\beta(y)\frac{\delta^0_\mu}{\mathbf{G}} \cr
ie\psi^\beta(y)\delta^\mu_\ell\frac{\mathbf{K}}{\mathbf{G}}\partial^\ell & -ie\psi^\dagger_\beta(y)\delta^\mu_\ell\frac{\mathbf{K}}{\mathbf{G}}\partial^\ell\cr
ie\psi^\beta(y)\frac{\partial_j}{\mathbf{G}} & - ie\psi^\dagger_\beta(y)\frac{\partial_j}{\mathbf{G}}\cr
0 & 0 \cr
0 & 0 \cr
ie\psi^\beta(y)\frac{M^2}{m_s\mathbf{G}} &  -ie\psi^\dagger_\beta(y)\frac{M^2}{m_s\mathbf{G}}\cr 
-ie\psi^\beta(y)\frac{1}{\mathbf{G}} & ie\psi^\dagger_\beta(y)\frac{1}{\mathbf{G}} \cr
 0 & 0 \cr
-ie\psi^\beta(y)\frac{1}{\mathbf{G}} & ie\psi^\dagger_\beta(y)\frac{1}{\mathbf{G}}\cr
 0 & 0 \cr
\end{pmatrix}\;,
\end{equation}
and
\begin{equation}
S = \begin{pmatrix}
-ie\psi^\alpha(x)\frac{\delta^0_\nu}{\mathbf{G}} & ie\psi^\alpha(x)\delta^\nu_\ell\frac{\mathbf{K}}{\mathbf{G}}\partial^\ell & ie\psi^\alpha(x)\frac{\partial_k}{\mathbf{G}} & 0 & 0 & -ie\psi^\alpha(x)\frac{M^2}{m_s\mathbf{G}} & ie\psi^\alpha(x)\frac{1}{\mathbf{G}} & 0 & ie\psi^\alpha(x)\frac{1}{\mathbf{G}} & 0 \cr
ie\psi^\dagger_\alpha(x)\frac{\delta^0_\nu}{\mathbf{G}} & -ie\psi^\dagger_\alpha(x)\delta^\nu_\ell\frac{\mathbf{K}}{\mathbf{G}}\partial^\ell & -ie\psi^\dagger_\alpha(x)\frac{\partial_k}{\mathbf{G}} & 0 & 0 & ie\psi^\dagger_\alpha(x)\frac{M^2}{m_s\mathbf{G}} & -ie\psi^\dagger_\alpha(x)\frac{1}{\mathbf{G}} & 0 & -ie\psi^\dagger_\alpha(x)\frac{1}{\mathbf{G}} & 0 
\end{pmatrix}\;.
\end{equation}
{with the same definitions $\mathbf{G}\overset{!}{=}M^2-\mathbf{K}\nabla^2_x$ and  $ \mathbf{K}\overset{!}{=}1+\dfrac{\square_x}{m^2}$ previously adopted for the free case discussion.}\\
\indent From this, we recognize the following non-null equal-time super Poisson brackets involving fermion fields
\begin{align}
\lbrace \psi^\alpha(x) ; \psi^\dagger_\beta(y)\rbrace & = -i\delta^\alpha_\beta \delta^3(\vec{x}-\vec{y})\quad,\\
\lbrace A_0(x);\psi^\beta(y)\rbrace = ie\psi^\beta(y)\mathbf{G}_x^{-1}\delta^3(\vec{x}-\vec{y})
\quad &\;\,,\quad \lbrace A_0(x);\psi^\dagger_\beta(y)\rbrace = -ie\psi^\dagger_\beta(y)\mathbf{G}_x^{-1}\delta^3(\vec{x}-\vec{y})\quad,\\
\lbrace\Pi^j(x);\psi^\beta(y)\rbrace=ie\psi^\beta(y)\dfrac{\mathbf{K}_x}{\mathbf{G}_x}\partial^j_x\delta^3(\vec{x}-\vec{y})\quad &\;\,,\quad\lbrace\Pi^j(x);\psi^\dagger_\beta(y)\rbrace=-ie\psi^\dagger_\beta(y)\dfrac{\mathbf{K}_x}{\mathbf{G}_x}\partial^j_x\delta^3(\vec{x}-\vec{y})\quad,\\
\lbrace\Gamma_j(x);\psi^\beta(y)\rbrace = ie\psi^\beta(y)\dfrac{\partial_j^x}{\mathbf{G}_x}\delta^3(\vec{x}-\vec{y})\quad&\;\,,\quad\lbrace\Gamma_j(x);\psi^\dagger_\beta(y)\rbrace = -ie\psi^\dagger_\beta(y)\dfrac{\partial_j^x}{\mathbf{G}_x}\delta^3(\vec{x}-\vec{y})\quad,\\
\lbrace\pi(x);\psi^\beta(y)\rbrace=ie\psi^\beta(y)\dfrac{M^2}{m_s\mathbf{G}_x}\delta^3(\vec{x}-\vec{y})\quad&\;\,,\quad\lbrace\pi(x);\psi^\dagger_\beta(y)\rbrace=-ie\psi^\dagger_\beta(y)\dfrac{M^2}{m_s\mathbf{G}_x}\delta^3(\vec{x}-\vec{y})\quad.
\end{align}
These brackets along the presented in \eqref{F.PB} constitute the complete set of classical brackets of the Podolsky-Stueckelberg electrodynamics. \\
In order to make the functional quantization of the interacting case, it is important to recall that the generating functional is given by \eqref{GBAppr}, but in this time considering the extra terms involving the fermionic fields\footnote{It is worth mentioning that from now on we will consider the replacement $\psi^\dagger_\alpha\to\bar{\psi}_\alpha$ (which Jacobian does not affect the functional measure), for the sake of convenience. On the other hand, since in this case $\zeta^{st}=\begin{pmatrix}
\zeta_F & \psi^\alpha & \bar{\psi}_\alpha\end{pmatrix}$, the functional measure now reads: $\mathscr{D}[\zeta]=\mathscr{D}[\zeta_F]\displaystyle\left(\prod_{\alpha=1}^4\mathscr{D}[\psi^\alpha]\right)\left(\prod_{\alpha=1}^4\mathscr{D}[\bar{\psi}_\alpha]\right)$.}
\begin{equation}
Z = N\int\exp\left\lbrace i\int \mathcal{L}^{(3)}_{dyn}d^4x\right\rbrace\left|\det\left(\omega^{(3)}\right)\right|^{1/2}\delta[\,\mathcal{G}\,]\delta[\mathcal{G}^{(1)}]\delta[\mathcal{G}^{(2)}_1]\delta[\mathcal{G}^{(2)}_2]\mathscr{D}[A_\mu , \Pi^\mu , \Gamma_j, \Phi^j, B , \pi , \psi^\alpha , \psi^\dagger_\alpha]\;,
\end{equation}
with
\begin{align}
\mathcal{L}^{(3)}_{dyn}=\mathcal{L}^{(3)}_{dyn,F}+\dfrac{i}{2}\left\lbrace\, \bar{\psi}_\alpha[\gamma^\mu]^\alpha{}_\beta\partial_\mu\psi^\beta-\partial_\mu\bar{\psi}_\alpha[\gamma^\mu]^\alpha{}_\beta\psi^\beta\,\right\rbrace-m_e\bar{\psi}_\alpha\psi^\alpha-eA_k\bar{\psi}_\alpha[\gamma^k]^\alpha{}_\beta\psi^\beta\;,
\end{align}
and
\begin{equation}
\det\left(\omega^{(3)}\right)=\det\left(\omega^{(3)}_F\right), \quad \mathcal{G}=\mathcal{G}_F, \quad \mathcal{G}^{(1)}=\mathcal{G}^{(1)}_F+e\psi^\dagger_\alpha\psi^\alpha, \quad \mathcal{G}^{(2)}_1=\mathcal{G}^{(1)}_{1,F}, \quad \mathcal{G}^{(2)}_2=\mathcal{G}^{(2)}_{2,F}-e\psi^\dagger_\alpha\psi^\alpha\;.
\end{equation}
The procedure is essentially the same as for the non-interacting case with few differences in the \eqref{Auxint1} and \eqref{GW1} steps, due to the presence of the term involving the $\psi-$fields. Thus, we get:
\begin{eqnarray}
&&Z=N\det(\mathbf{G})\int\exp\left\lbrace i\int\left[-\dfrac{1}{4}F_{\mu\nu}F^{\mu\nu}+\dfrac{1}{2m^2}\partial_\mu F^{\mu\rho}\partial^\nu F_{\nu\rho}+\dfrac{M^2}{2}\left(A_\mu+\partial_\mu\dfrac{B}{m_s}\right)\left(A^\mu+\partial^\mu\dfrac{B}{m_s}\right)+ \right.\right.\cr\cr
&&\left.\left. \qquad
+\dfrac{i}{2}\Big(\bar{\psi}_\alpha[\gamma^\mu]^\alpha{}_\beta\partial_\mu\psi^\beta-\partial_\mu\bar{\psi}_\alpha[\gamma^\mu]^\alpha{}_\beta\psi^\beta\Big)-m_e\bar{\psi}_\alpha\psi^\alpha-eA_\mu\bar{\psi}_\alpha[\gamma^\mu]^\alpha{}_\beta\psi^\beta\right]d^4x\right\rbrace\times \cr\cr
&&
\qquad\hspace{8.7cm}\times\delta\left[\mathcal{G}^{(2)}_1\right]\mathscr{D}[A_\mu, B, \psi^\alpha , \bar{\psi}_\alpha]\cr\cr
&&\equiv N\det(\mathbf{G})\int\exp\Big\lbrace i\,S[A_\mu, B, \psi^\alpha , \bar{\psi}_\alpha]\Big\rbrace\delta\left[\left(1+\dfrac{\square}{m^2}\right)\partial^jA_j-\dfrac{M^2}{m_s}B\right]\mathscr{D}[A_\mu, B, \psi^\alpha , \bar{\psi}_\alpha]\;.
\end{eqnarray}
As in \eqref{noncov.Z.free}, the expression obtained for the generating functional is non-covariant and so we could proceed as we did in the free case to obtain the desired manifestly covariant expression. However, a more elegant covariant expression for $Z$ compatible with the Coulomb-Podoslky-Stueckelberg gauge is a sort of generalization of the so-called $R_\epsilon-$gauges, whose construction involves a family of Lorenz-Podolsky-Stueckelberg gauges-like characterized by a scalar function $f$, and multiplied by a suitable Gaussian weight parameterized by $\epsilon$ \cite{m2y,fix1,fix2}.
\begin{itemize}
    \item[$(i).$] Firstly, according to Faddeev and Popov, we introduce the following object (proportional to identity due to a factor depending on $\epsilon$) into the expression for $Z$
    \begin{eqnarray}
        1\;\;\sim&& \det\left[\left(1+\dfrac{\square}{m^2}\right)\square+M^2\right] \int \delta\left[\left(1+\dfrac{\square}{m^2}\right)\partial^\mu A^\Lambda_\mu-\dfrac{M^2}{m_s}B^\Lambda-f\right]\times\cr\cr
        &&\hspace{4cm}\times\exp\left\lbrace -\dfrac{i}{2\epsilon}\int f^2(x) d^4x\right\rbrace \mathscr{D}[f]\mathscr{D}[\Lambda]\;.
    \end{eqnarray}
    \item[$(ii).$]By performing a reverse gauge transformation, we get\footnote{Here again the normalization factor contains the (infinite) gauge volume group.}
    \begin{eqnarray}
        Z&=& N\det(\mathbf{G})\det(\mathbf{K}\square+M^2)\int \exp\Big\lbrace i\,S[A_\mu, B, \psi^\alpha , \bar{\psi}_\alpha]\Big\rbrace\times \cr\cr
        &&\times \delta\left[\left(1+\dfrac{\square}{m^2}\right)\partial^\mu A_\mu-\dfrac{M^2}{m_s}B-f\right]\delta\left[\mathbf{G}(\Lambda)+\left(1+\dfrac{\square}{m^2}\right)\partial^jA_j-\dfrac{M^2}{m_s}B\right]\times\cr\cr
        &&\hspace{4.1cm}\times\exp\left\lbrace -\dfrac{i}{2\epsilon}\int f^2(x) d^4x\right\rbrace \mathscr{D}[f]\mathscr{D}[\Lambda]\mathscr{D}[A_\mu, B, \psi^\alpha , \bar{\psi}_\alpha]\nonumber \\
    \end{eqnarray}
    Reducing the second delta functional to $\delta\big[\Lambda+\mathbf{G}^{-1}(\mathbf{K}\partial^jA_j-\frac{M^2}{m_s}B)\big]/\det(\mathbf{G})$ and integrating over $\Lambda$, we get
\begin{eqnarray}
        Z&=& N\det(\mathbf{K}\square+M^2) \int \exp\Big\lbrace i\,S[A_\mu, B, \psi^\alpha , \bar{\psi}_\alpha]\Big\rbrace\times \cr\cr
        &&\times \delta\left[\left(1+\dfrac{\square}{m^2}\right)\partial^\mu A_\mu-\dfrac{M^2}{m_s}B-f\right]\exp\left\lbrace -\dfrac{i}{2\epsilon}\int f^2(x) d^4x\right\rbrace \mathscr{D}[f]\mathscr{D}[A_\mu, B, \psi^\alpha , \bar{\psi}_\alpha]\nonumber \\
    \end{eqnarray}
\end{itemize} 
Therefore, we end up with the final expression for $Z$
\begin{multline} 
Z=N\det(\mathbf{K}\square+M^2) \int \exp\left\lbrace -\frac{i}{2\epsilon}\int\left[\Big(1+\frac{\Box}{m^2} \Big)\partial^{\mu} A_\mu-\frac{M^2}{m_s}B\right]^2d^4x\right\rbrace\times
 \\
 \times\exp\Big\lbrace i\,S[A_\mu, B, \psi^\alpha , \bar{\psi}_\alpha]\Big\rbrace\mathscr{D}[A_\mu, B, \psi^\alpha , \bar{\psi}_\alpha]\;
 \label{Z}
 \end{multline}
 In this Abelian theory case, the only remaining operator normalization can be written in terms of a Faddeev-Popov ghost pair with no coupling with the physical fields. Although it is not important for the present zero temperature case, this indeed has implications in the thermodynamics of the theory. Regarding the degrees of freedom, our final phase superspace has, separating bosonic and fermionic excitations, $20+2\times4$ dimensions. However, eliminating the four local excitations associated with the auxiliary fields, and considering the four constraints between the phase space fields, there are $12+2\times 4$ remaining independent quantities. It corresponds to $6+4$ configuration space local excitations. These are exactly the ones that correspond to two massive spin $1$ fields corresponding to the poles and one Dirac fermion with its $4$ degrees of freedom. It is worth mentioning that it reinforces the pure gauge character of the auxiliary Stueckelberg field. Despite being just an artifice, this auxiliary field turns the system into a gauge invariant one, opening the possibility of deriving Ward-Takahashi non-perturbative constraints even for the Boson mass renormalization constant.

\section{Conclusion and final remarks}
\label{sc8}

As it was possible to testify in the text, we fully investigated the quantization of the GS model in the path integral formalism in the Faddeev-Jackiw framework. First, we present some classical, quantum and thermal physics issues associated with the studied model. Continuing on, we present the standard formulation of FJ in the language of differential forms as well as exterior derivatives and instruct how to deal with the constraints that arise throughout the algorithm. Then we apply the FJ program to analyze the free case of the GS model, deriving the entire symplectic structure and obtaining the Poisson brackets of the model. The path integral quantization of the free GS model was made by implementing the correct functional measure of integration taking into account the Darboux theorem. Finally, after extending the FJ theory for Grassmann-valued fields, not only the complete FJ setup could be obtained for the GS electrodynamics (interacting case) but also the correct integration measure, which was implemented in the path integral quantization, following the same steps as the free case. 

Regarding the future perspectives, the formal structure provided in this work can be applied to a wide class of systems: the Gross-Neveau self-interacting fermionic model \cite{Gross}, on the superconductivity and some other correlated descriptions of condensed matter. Moreover, it is also suitable to be applied in the context of supersymmetry \cite{Lopez} and bosonization procedures \cite{Mantilla}. Since our approach is capable to characterize non-linearly interacting systems, the study of non-Abelian BF models \cite{Mansfield} can be conveniently performed since it is already a first-order theory. Another perspective is the study of the renormalization and radiative corrections for the interacting generalized Stueckelberg model. The calculation of the anomalous magnetic moment and comparison with the well-established experimental data can be a method to constrain the Proca and Podolsky masses\footnote{There is a possibility to derive different bounds, since the known ones are related to models in which one of these
constants is absent.}. Moreover, determining its associated partition function possibly leads to complementary constraints for the system’s parameters due to thermodynamic phenomenology.

\section*{Acknowledgement}
LGC thanks Coordination for the Improvement of Higher Education (CAPES) for financial support. GBG thanks to The São Paulo Research Foundation (FAPESP)  Post Doctoral grant No. 2021/12126-5. AAN thanks Federal University of Alfenas (UNIFAL) for its hospitality in his temporary stay as a visiting professor. BMP thanks National Council for Scientific and Technological Development (CNPq) for partial financial support.

\end{document}